\documentclass[aip,sd,amsmath,amssymb,preprint,xcolor]{revtex4-1} 
\usepackage{amsmath}
\usepackage{paralist}
\usepackage{graphicx}
\usepackage{chemfig}
\usepackage{rotating}
\usepackage{longtable}
\usepackage{booktabs}

\usepackage{accents}
\newcommand{\ubar}[1]{\underaccent{\bar}{#1}}

\usepackage{xcolor}
\usepackage{ulem,cancel}

\newcommand{\onlysi}[1]{}
\newcommand{\rcm}{cm$^{-1}$}

\renewcommand{\emph}[1]{{\it #1}}
\newcommand{\mbf}[1]{{\mathbf #1}}

\newcommand{\beq}{\begin{equation}}
\newcommand{\eeq}{\end{equation}}

\newcommand{\chan}[1]{{\color{black!40!red!100}{#1}}}
\renewcommand{\chan}[1]{#1}

\begin{document}
\title{
Linear-scaling generation of potential energy surfaces using a double incremental expansion
}
\author{Carolin K\"onig}
\email{carolink@kth.se}
\altaffiliation{Present address: KTH Royal Institute of Technology, 
School of Biotechnology, Division of Theoretical Chemistry and Biology, S-106 91 Stockholm, Sweden }
\author{Ove Christiansen}
\email{ove@chem.au.dk}
\affiliation{
Department of Chemistry, Aarhus University, DK-8000 Aarhus C, Denmark.
}
\date{\today}

\begin{abstract}

We present a combination of the incremental 
 expansion of potential energy surfaces (PESs), known as $n$-mode expansion,
  with the incremental 
 evaluation of the electronic energy in a many-body approach. 
The application of semi-local coordinates in this context 
 allows the generation of 
 PESs in a very cost-efficient way.
For this, we employ the recently introduced flexible adaptation of local coordinates 
 of nuclei (FALCON) coordinates.
By introducing an additional transformation step, 
  concerning only a fraction of the vibrational degrees of freedom, we can  
 achieve linear scaling of \chan{the {\it accumulated} cost} of the  
 single point calculations required in the PES generation. 
Numerical examples of these double incremental  approaches for 
  oligo-phenyl examples show fast convergence
 with respect to the maximum number of 
 simultaneously treated fragments and only a modest error introduced by the additional 
 transformation step.
The approach, presented here, represents a major step towards the applicability of vibrational
 wave function methods to sizable, covalently bound systems.

\end{abstract}

\maketitle
\section{Introduction \label{Intro}}

Potential energy surfaces (PESs) are essential ingredients for quantum-dynamical 
 calculations used for the simulation of vibrational spectra and 
 chemical reactions. 
The focus of the present work is the construction of PESs expanded around a reference point 
for the use in subsequent anharmonic vibrational wave function calculations. 
The computational cost of generating full PESs is determined by (i) the number of required
 single point calculations (SPCs) and (ii) the computational cost per SPC.
In a full-dimensional PES generation on a grid, the number of required SPCs 
 increases 
 exponentially with the number of dimensions and the computational cost per SPC 
 scales according to the applied electronic structure method.
A full PES generation is, hence, only affordable for 
 molecular systems of a few atoms.
A common approximation to overcome the exponential scaling in the number of SPCs 
 is to restrict the direct mode--mode couplings\cite{Jung96,roit1997,cart1997,chab1999,yagi2000}   
 in an incremental-type of expansion.
Similarly, the scaling of the electronic structure calculations can 
 be reduced \chan{to 
 linear} scaling by incremental 
 fragmentation approaches when exploiting the locality of the 
 interaction (\chan{see, e.g., Ref.~\onlinecite{deev2005}}).
Both of these approaches rely on an incremental expansion and are 
 approximate, but have proven cost effective.
In this work, we show, that these two approaches can beneficially be combined. 
A detailed analysis shows that 
 the computational gains are particularly large when (semi-)local coordinates are employed.
By \emph{semi-local} we \chan{describe} that the majority of the coordinates 
 are strictly local to one or a few fragments.
We can even obtain \emph{linear} scaling
 of the accumulated computational cost of all SPCs required for the 
 PES  generation
 with increasing molecular size. 
This can be achieved by 
 employing auxiliary \chan{vibrational coordinates} for certain vibrational degrees of freedom
 and requires the introduction of an additional transformation step for 
 certain contributions to the PES representation. 
Notable, this scaling considers the number of SPCs 
 as well as the computational cost of the individual SPCs,
 and is to our knowledge the first report of such low scaling for anharmonic PES construction
 considering the full vibrational space. 

The \chan{above-mentioned} restriction of direct mode--mode couplings\cite{Jung96,roit1997,cart1997,chab1999,yagi2000}
 includes the special case of the pair approximation\cite{Jung96} 
 and is known under different names, such as many-body expansion\cite{scha1989} 
 $n$-mode representation\cite{cart1997}, 
 mode-coupling expansion, cut-HDMR (High Dimensional Model Representation)\cite{rabi1999}, 
 cluster expansion\cite{meye2012}, or others\cite{grie2006}.
We will generally refer to this expansion as $n$-mode representation or $n$-mode expansion. 
This approach reduces the scaling in required grid points for 
 a PES generation with increasing number of modes  
 from exponential scaling to  polynomial scaling with 
 the $n$th power, where $n$ is the maximum number of simultaneously considered modes. 
Still, this scaling prevents its application to systems larger than a few tens of vibrational
 degrees of freedom, 
 especially when combined with decent (and thereby computationally expensive) electronic
 structure methods.
Accordingly, the $n$-mode representation has been combined with a number 
 of approaches further reducing the number of required SPCs, 
 such as screening schemes\cite{beno2004,rauh2004,beno2006,pele2008,beno2008,seid2009b} and adaptive choice of the 
 grid points\cite{spar2009,rich2012,stro2014}.
Another possibility to reduce the computational scaling for PES generations is  
 to obtain the expensive higher-order mode couplings in a more approximate manner\cite{rauh2004,yagi2007,rauh2008,spar2009b,spar2010,meie2013}.
This can, for example, be achieved by using lower-cost electronic structure methods\cite{rauh2004,yagi2007,rauh2008,spar2009b},
  derivative information\cite{spar2010}, or other approximate ways to calculate the 
 individual single point energies\cite{meie2013}.
In special cases, effective linear scaling of the number of required SPCs has been observed
 with an adaptive grid approach\cite{hans2010}. 
Both screening as well as adapted choice of required grid points are expected 
 to be particularly beneficial, when combined with adapted coordinates with some kind of spatial locality
\cite{pane2014,chen2014,rich2014,pane2016}.
Still, even if linear scaling of the number of grid points can be achieved 
(which is likely
 to be  
 system and coordinate dependent for the described approaches), the computational cost 
 of every grid point still scales according to the chosen electronic structure method
 with system size.
This typically leads to overall super-linear scaling with system size. 
 
Incremental and fragmentation ideas for electronic calculations 
 have been found in the literature for several decades now 
  and have raised considerable attention within the last decades 
 (see the historical review in Ref.~\onlinecite{coll2015}).
This is reflected in a large number of different incremental 
 fragmentation approaches \cite{fuji1981,kita1999,zhan2003,deev2005,huan2005,li2007}.
Examples are the fragment molecular orbital (FMO)\cite{fuji1981,kita1999,fedo2006} model, 
 the molecular 
 fractioning with conjugated caps (MFCC) scheme\cite{zhan2003}, the systematic molecular fragmentation 
 (SMF) method \cite{deev2005}, and the generalized energy-based fragmentation (GEFB)
  method\cite{li2007}, just to name a few.
The fragmentation idea has also  been combined with multi-layer methods.\cite{fedo2005,mei2006,dahl2007,mull2009,bera2009,mayh2011,reza2010}
A full account of all fragmentation methods is far beyond the scope of the present 
 introduction. 
We refer to a recent review focusing on FMO methods \cite{gord2012} as well 
 as comprehensive reviews \chan{on energy-based fragmentation methods}\cite{coll2015,ragh2015}
 and other recent attempts to classify the different methods \cite{mayh2011} and/or 
 generalize the underlying expressions and ideas to a unified framework \cite{rich2012b,mayh2012,rich2013}.

Our double incremental approach 
 concerns a combination of  
 the incremental expansion of the PES  in terms of vibrational degrees of 
 freedom \chan{(known as $n$-mode representation)}
 with that of the electronic energy in terms of fragments in a flexible manner. 
We are not aware of any previous work achieving such a flexible setup.
This makes it  fundamentally different from the \emph{many-body expansion} approach for PES
 generation of Varandas and Murrel\cite{vara1977} and its so-called 
 \emph{double many-body expansion}\cite{vara1984} extension, despite the similar names.
In the present work, particular gains are obtained by employing fairly local 
 \chan{vibrational} coordinates.
The attractiveness of employing "local" modes in combination with the $n$-mode representation
is also exploited in the local monomer model by Bowman and co-workers \cite{wang2010,wang2012,bowm2015}. 
Indeed, one of the main characteristics of the local monomer model is that 
 all coordinates are fully localized to a particular fragment and as such it has been very 
 successfully applied in the context of PES generation for molecular clusters. 
Our setup is not restricted to local vibrational coordinates only.
The presented double incremental scheme can thereby be applied to all types of 
 \chan{vibrational} coordinates and 
particular savings are obtained for all,
 though significant savings are most easily achievable for coordinates that are strictly local to part of the system.
Our methods do, however, not require that all coordinates are local.
Thereby our approach is applicable to covalently bound systems.
Still, to achieve a significant reduction of the computational cost of the 
 presented double incremental approach
 compared to conventional approaches, semi-local coordinates should be applied. 
In the present work, we make use of the recently introduced FALCON  (Flexible Adaptation of Local COordinates of Nuclei) scheme\cite{koni2016},
which aims at constructing exactly such coordinate sets.  
The resulting rectilinear FALCON coordinates are constructed in a way that
  a large fraction of the vibrational coordinates are
 local to certain groups of atoms (\chan{i.e., local to fragments}) 
 and therefore enable a spatially 
 motivated incremental expansion of the PES. 
Using these coordinates in our double incremental expansion will be denoted \emph{double 
 incremental expansion in FALCON coordinates} (DIF). 
In \chan{this model}, the part of the FALCON coordinates that are not fully local 
 can lead to a super-linear scaling of the number of required SPCs. 
This is why, we have devised an additional protocol, which generates local auxiliary coordinates 
 covering the same vibrational space as the common FALCON coordinates for a local 
 combination of fragments. 
\chan{These auxiliary coordinates are generally different for different 
 fragments and combinations of fragments.
 Performing the SPCs for the PES construction along}
 these auxiliary coordinates 
 allows linear scaling in the accumulated cost of SPCs. 
It, however, requires the introduction of an additional transformation step 
 for the \chan{local PES representation to be used in an overall PES expansion}.
We term this approach \emph{double incremental expansion \chan{in FALCON coordinates}
 with auxiliary coordinate transformation} (DIFACT).
This transformation of parts of the PES is, however, only exact in the 
limit of a complete PES and therefore introduces an additional potential source of error
 in practical calculations. 
This lifts 
the exact limits of the incremental expansion in case of DIFACT and must be 
 numerically validated. 

In this article, we derive and motivate the DIF as well as the DIFACT model starting 
from general incremental expansion including scaling considerations (Section \ref{sec:theo}). 
We furthermore present 
 a pilot implementation (Section \ref{sec:impl}), computational setup
 (Section \ref{sec:cd}) as well as first numerical results for tetra- and hexa-phenyl 
 (Section~\ref{sec:tech_test}).
After concluding from our results, we give a perspective on future 
 extensions and applications of the presented method (Section \ref{sec:conc}).  

\section{Theory \label{sec:theo}}

\subsection{General incremental expansion}

In this section, we will set up a common nomenclature for 
 incremental expansions, which is capable of describing the 
 $n$-mode expansion of PESs and the incremental 
 expansion of the energy (often denoted by many-body \chan{expansion} \cite{xant1994})
 at a given point using the same principles and nomenclature.
As outlined in the Introduction both expansions are well-known in literature
 and the purpose of the present section is to reformulate the basic ideas
 in a general and convenient form. Thus, deviations from standard formulations found in literature
 are not fundamental, but necessary to formulate the double incremental expansion
 in a compact and general manner in Sections II~B -- II~G. 

The main idea of \chan{incremental} expansions is that a multi-dimensional 
 function \linebreak
 $F(x_1, x_2, \dots, x_n)$ can be expanded around a reference point $F^0= F(0,0, \dots, 0)$ with indices running from $1$ to $n$ as
\begin{align}  
F(x_1, x_2, \dots, x_n) 
 = & 
F^{0}
+ \sum_{i=1}^n [F^{\rm 1D}_i(x_i) -F^0] \notag \\
&
+ \sum_{j>i=1}^n \left\{ [F_{ij}^{\rm 2D}(x_i,x_j) -F^0] - [ F_i^{\rm 1D}(x_i) -F^0]  - 
 [ F_j^{\rm 1D}(x_j) - F^0 ]\right\}
 \notag \\
& + \sum_{k>j>i=1}^n \left\{ [F^{\rm 3D}_{ijk}(x_i,x_j,x_k)  -F^0]   \right. \notag  \\
& \qquad 
 - \left[ [F_{ij}^{\rm 2D}(x_i,x_j)-F^0]  - [ F_i^{\rm 1D}(x_i)-F^0]  - [ F_j^{\rm 1D}(x_j)- F^0] \right]  \notag\\
& \qquad- \left[ [F_{ik}^{\rm 2D}(x_i,x_k)-F^0]  - [ F_i^{\rm 1D}(x_i)-F^0]  - [ F_k^{\rm 1D}(x_k) -F^0] \right]  \notag\\
& \qquad - \left[[ F_{jk}^{\rm 2D}(x_j,x_k)-F^0]  - [ F_j^{\rm 1D}(x_j)-F^0]  -  [F_k^{\rm 1D}(x_k) -F^0] \right]  \notag\\
& \left.
 \qquad -   [F_i^{\rm 1D}(x_i)-F^0]  - [ F_j^{\rm 1D}(x_j)-F^0]  - [ F_k^{\rm 1D}(x_k)
 - F^0]
\right\} \notag \\
& + \dots,
\label{eq:incr_basic}
\end{align}
where the following lower-order \emph{cut functions} are used, 
\begin{align}  
F^{\rm 0} & \equiv F(0,0,\dots,0,0,0,\dots,0,0,0,\dots,0,0,0,\dots,0)\\
F^{\rm 1D}_{i}(x_i) & \equiv F(0,0,\dots,0,x_i,0,\dots,0,0,0,\dots,0,0,0,\dots,0)\\
F^{\rm 2D}_{ij}(x_i,x_j) &  \equiv F(0,0,\dots,0,x_i,0,\dots,0,x_j,0,\dots,0,0,0,\dots,0) \\
F^{\rm 3D}_{ijk}(x_i,x_j,x_k) &  \equiv F(0,0,\dots,0,x_i,0,\dots,0,x_j,0,\dots,0,x_k,0,\dots,0).
\end{align}  
This expansion is exact in the limit where the order of the expansion is equal to the number of
 variables.

Introducing the \chan{bar functions},
\begin{align}
\bar{F}^{\rm 0D} &   \equiv F^0\\
\bar{F}^{\rm 1D}_i(x_i) &   \equiv {F}^{\rm 1D}_i(x_i) - \bar{F}^{\rm 0D}\\
\bar{F}^{\rm 2D}_{ij}(x_i,x_j) 
 &   \equiv  {F}^{\rm 2D}_{ij}(x_i,x_j) - \bar{F}^{\rm 1D}_i(x_i) - \bar{F}^{\rm 1D}_j(x_j)  - \bar{F}^{\rm 0D} \\
\bar{F}^{\rm 3D}_{ijk}(x_i,x_j,x_k) & \equiv {F}^{3D}_{ijk}(x_i,x_j,x_k) 
 - \bar{F}^{2D}_{ij}(x_i,x_j) - \bar{F}^{\rm 2D}_{ik}(x_i,x_k)  - \bar{F}^{\rm 2D}_{jk}(x_j,x_k) \notag \\
& \qquad -  \bar{F}^{\rm 1D}_i(x_i) - \bar{F}^{\rm 1D}_j(x_j) - \bar{F}^{\rm 1D}_k(x_k)   
 - \bar{F}^{\rm 0D}\\
& \vdots  \notag 
\end{align}
we can rewrite Eq.~(\ref{eq:incr_basic}) as
\begin{align}  
F(x_1, x_2, \dots, x_n) 
 = & 
\bar{F}^{\rm 0D}
+ \sum_{i=1}^n \bar{F}^{\rm 1D}_i(x_i)
+ \sum_{j>i=1}^n \bar{F}^{\rm 2D}_{ij}(x_i,x_j) 
+ \sum_{k>j>i=1}^n \bar{F}^{\rm 3D}_{ijk}(x_i,x_j,x_k) 
+ \dots
.
\label{eq:incr_bar}
\end{align}
In analogy to the set logic applied in Ref.~\onlinecite{kong2006},
 we define a variable combination range (VCR) containing 
 all variable combinations that are explicitly parameterized in the expansion.  
 Variable combinations are simply sets of variables
 and denoted in bold, such as ${\mbf c}_l$. 
It is often convenient to include
 $l$ in the notation, where $l$ is the number of variables in this set. 
The VCR is required to be \emph{closed on forming subsets}. 
A VCR is \emph{closed on forming subsets}, if for each variable combination 
 ${\mbf c}_l$ in the
 VCR, also all subsets of ${\mbf c}_l$ are considered.
This definition includes the empty set in all cases.   
\chan{For instance, for  a three-variable function $F(i,j,k)$} that is treated to second order in the 
 incremental expansion, the variable combination range is 
 $\{\{\},\{i\},\{j\},\{k\},\{i,j\},\{i,k\},\{j,k\}\}$.  

We can now rewrite Eq.~(\ref{eq:incr_bar}) in a compact and 
 flexible form as
\begin{align}  
F(x_1, x_2, \dots, x_n) 
 = & 
\sum_{{\mbf c}_l \in \{{\rm VCR}\}} \bar{F}^{{\mbf c}_l}(\{x \}^{{\mbf c}_l})
,
\label{eq:general_incr_expansion}
\end{align}
where the equal sign only holds for a complete VCR.
 $\bar{F}^{{\mbf c}_l}(\{x \}^{{\mbf c}_l})$ is the \chan{bar function} for 
 the variable combination ${{\mbf c}_l}$, i.e., a function of $l$ variables.
 The particular set of variables given by ${{\mbf c}_l}$ \chan{is} extracted from
 the full set, as indicated by the notation $\{x \}^{{\mbf c}_l}$. 
For instance for ${{\mbf c}_l} = {\mbf c}_3 =\{i,j,k\}$, 
 we have $\bar{F}^{\{i,j,k\}}(\{x \}^{\{i,j,k\}}) = \bar{F}^{\rm 3D}_{ijk}(x_i,x_j,x_k)$.

Using the introduced notation we express the \chan{bar potentials} generally as
\begin{align}
\bar{F}^{{\mbf c}_l} (\{x \}^{{\mbf c}_l})
 = {F}^{{\mbf c}_l}(\{x \}^{{\mbf c}_l}) - 
\sum_{
\substack{
{\mbf c}_{s} \subset {\mbf c}_{l}\\
{\mbf c}_{s} \in \{{\rm VCR}\}
}
}
\bar{F}^{{\mbf c}_s}(\{x \}^{{\mbf c}_s}).
\label{eq:incr_bar_combi}
\end{align}
This formulation not only allows for more compact expressions, but also enables 
 a more flexible parameterization of the expansion, such as 
    opening for different treatment of variables with different degree of correlation.
We accordingly restrict the sum in Eq.~(\ref{eq:incr_bar_combi}) to those lower-order 
 fragment combinations present in the VCR.
This restriction is not needed whenever the VCR is closed on forming subsets (as assumed by default).
We note, that with this restriction,  Eq.~(\ref{eq:incr_bar_combi}) also holds in 
 case of effective VCRs\chan{, which are introduced} in Section~\ref{sec:eff_vcr}.

In the following, \chan{we need} to express the bar functions for a variable combination 
  in terms of the original, uncorrected contributions, i.e., cut functions. 
Most generally, we can write
\chan{
\begin{align}
\bar{F}^{{\mbf c}_l} (\{x \}^{{\mbf c}_l})= 
\sum_{\substack{
{\mbf c}_{l'} \subseteq {\mbf c}_{l}\\
{\mbf c}_{l'} \in \{{\rm VCR}\}
}}
k^{{\mbf c}_{l'},{\mbf c}_l} 
{F}^{{\mbf c}_{l'}} (\{x \}^{{\mbf c}_{l'}}),
\label{eq:incr_bar_from_contr}
\end{align}
}
where $k^{{\mbf c}_{l'},{\mbf c}_{l}}$ is the coefficient of 
 ${F}^{{\mbf c}_{l'}}(\{x \}^{{\mbf c}_{l'}})$ 
 in $\bar{F}^{{\mbf c}_l}(\{x \}^{{\mbf c}_l})$.
\chan{As a consequence of the closed under forming subset condition of the VCR,
 all lower-order variable combinations are contained in the VCR.}
In this case, the coefficient amounts to $k^{{\mbf c}_{l'},{\mbf c}_l} =  (-1) ^{l-l'}$ 
 (see also Ref.~\onlinecite{kong2006}).

\subsection{Effective variable combination ranges \label{sec:eff_vcr}}

As stated before, we generally require the VCR to be closed on forming subsets.
Depending on the chosen VCR, it can, however, happen that the coefficients of the
 cut functions for some variable combinations are zero in the overall expansion.
\chan{Accordingly, these variable combinations are omitted in the corresponding 
 \emph{effective} VCR.}
These cases occur in VCRs in which not all 
 variable couplings are included up to the same coupling level. 
To identify these terms, we first write the full function as a weighted sum of its 
 cut functions, i.e., we combine Eqs.~(\ref{eq:general_incr_expansion}) 
 and (\ref{eq:incr_bar_from_contr}) to 
\begin{align}
F(\{x\}) 
= \sum_{{\mbf c}_{l} \in \chan{\{{\rm VCR}\}}} 
\sum_{\substack{
{\mbf c}_{l'} \subseteq {\mbf c}_{l}\\
\chan{
{\mbf c}_{l'} \in \{{\rm VCR}\}
}}}
 (-1)^{l-l'}{F}^{{\mbf c}_{l'}} (\{x\}^{{\mbf c}_{l'}} ) 
= \sum_{{\mbf c}_{l'} \in \chan{\{{\rm VCR}\}}} p_{{\mbf c}_{l'}}^{\rm VCR} {F}^{{\mbf c}_{l'}} (\{x\}^{{\mbf c}_{l'}} ) 
\label{eq:overall_exp}
\end{align}
with
\begin{align}
p_{{\mbf c}_{l'}}^{\rm VCR}=
\sum_{{\mbf c}_l \in {\rm VCR}; \chan{{\mbf c}_{l} \supseteq {\mbf c}_{l'}}} 
(-1)^{l-l'}
=\sum_{l=l'}^{l_{\rm max}} 
N_{{\mbf c}_{l'}, l}^{\rm VCR}
(-1)^{l-l'},
\label{eq:coef}
\end{align}
where $l_{\rm max}$ is the largest \chan{variable-combination order in the} VCR 
 and $N_{{\mbf c}_{l'}, l}^{\rm VCR}$ is the number of variable combinations of the 
 order $l$ that \chan{are supersets of  or equal to ${\mbf c}_{l'}$ and 
 are contained in the VCR,
 which is closed on forming subsets.}
\chan{For a more rigorous derivation of Eq.~(\ref{eq:coef}) we refer to Appendix~\ref{sec:app_A}.} 
We can, hence, use Eq.~(\ref{eq:coef}) to analyze the incremental expansion for a given VCR 
 and set up an \emph{effective} VCR, in which all variable combinations with zero contributions
 are omitted. 
\chan{Effective VCRs are} not generally closed on forming subsets. 
An incremental expansion with such an effective VCR, will, however, give the same results as
 the analogous expansion in the corresponding VCR that contains all subsets of variable combinations.
When setting up an effective VCR, special care should be taken, keeping in mind that the extension 
 of the VCR may lead to other previously omitted terms to have a non-zero contribution.

\chan{
Assume, we expand the function $F(x_A,x_B,x_C,x_D)$ in a variable combination range 
 ${\rm VCR_1} = \{\{\}, \{A\}, \{B\}, \{C\}, \{D\},  \{A,B\}, \{C,D\} \}$ which is closed on 
 forming subsets. 
We can calculate the coefficient  
 for the variable combination $\{A\}$ in an incremental expansion for ${\rm VCR_1}$ by 
\begin{align}
 p_{\{A\}}^{\rm VCR_1} & = 
 N_{\{A\},1}^{\rm VCR_1} (-1)^{1-\dim(\{A\})}
  +
 N_{\{A\},2}^{\rm VCR_1} (-1)^{2-\dim(\{A\})}, 
\label{eq:ex1}
\end{align}
where $N_{\{A\},1}^{\rm VCR_1} = 1 $ is the number of variable combinations 
 in ${\rm VCR_1}$ of order $1$ that are supersets of or equal to $\{A\}$.
This only holds for the variable combination $\{A\}$ itself. 
Similarly, for $N_{\{A\},2}^{\rm VCR_1} $, we count the number of variable combinations
 in ${\rm VCR_1}$ of order $2$ that are supersets of or equal to $\{A\}$.
This  only holds true 
 for $\{A,B\}$, so that $N_{\{A\},2}^{\rm VCR_1} = 1 $.
Inserting $N_{\{A\},1}^{\rm VCR_1}=N_{\{A\},2}^{\rm VCR_1} = 1 $ into Eq.~(\ref{eq:ex1}), 
 we obtain $p_{\{A\}}^{\rm VCR_1}=0$.
The corresponding coefficients for $\{B\}$, $\{C\}$, and $\{D\}$ are obtained in 
 complete analogy to that for $\{A\}$ for the given ${\rm VCR}_1$.
This means that the variable combinations  $\{A\}$, $\{B\}$, $\{C\}$, and $\{D\}$ 
 have zero contribution in the incremental expansion for ${\rm VCR}_1$.
We can, hence, 
  construct an effective VCR for ${\rm VCR}_1$ that contains only 
 variable combinations with non-zero coefficients $p_{{\mbf c}_{l'}}^{{\rm VCR}_1}$.
It reads $ {\rm VCR_1^{eff}} = \{\{\}, \{A,B\}, \{C,D\} \}$. 

Let us now consider the very similar 
 ${\rm VCR_2} = \{\{\}, \{A\}, \{B\}, \{C\}, \{D\},  \{A,B\}, \{A,C\},$ $\{C,D\} \}$. 
It differs from ${\rm VCR_1}$ only by the addition the variable combination $ \{A,C\}$.
The individual coefficients can be calculated with an equation that is similar to Eq.~(\ref{eq:ex1}), 
 but replaces ${\rm VCR_1}$ by ${\rm VCR_2}$.
Counting the variable combinations in  ${\rm VCR_2}$ that are supersets of or equal to  $\{A\}$ in
 the variable combination orders $1$ and $2$, separately, 
we obtain $N_{\{A\},1}^{\rm VCR_2} =1 $ and $N_{\{A\},2}^{\rm VCR_2} = 2 $.
The two variable combinations of order 2
 in ${\rm VCR_2}$ that contain  $\{A\}$
 are $\{A,B\}$ and $\{A,C\}$.
Employing  ${\rm VCR_2}$, the variable combination $\{A\}$ has, hence, the coefficient 
 $p_{\{A\}}^{\rm VCR_2}=-1$.
The same coefficient is obtained for $\{C\}$. 
The number of variable combinations of order 2
 in ${\rm VCR_2}$ that are supersets of $\{B\}$ is $N_{\{B\},2}^{\rm VCR_2} = 1 $.
Together with $N_{\{B\},1}^{\rm VCR_2} = 1 $, we obtain $p_{\{B\}}^{\rm VCR_2}=0$.
$p_{\{D\}}^{\rm VCR_2}=0$ 
 is obtained completely analogously. 
This means, that for ${\rm VCR_2}$, only $\{B\}$ and $ \{D\}$ and also 
 the empty set $\{\}$ (not shown) have zero contribution. 
With these coefficients we can now assemble an effective VCR for ${\rm VCR_2}$
 with only non-zero contributions.
It reads 
  ${\rm VCR_2^{eff}} =  \{\{A\},  \{C\}, \{A,B\}, \{A,C\}, \{C,D\} \}$.
Obviously, the variable combinations $\{A\}$ and $\{C\}$, which
 can be omitted in ${\rm VCR_1^{eff}}$, cannot be omitted in ${\rm VCR_2^{eff}}$, even though
 the only difference between ${\rm VCR_1}$ and ${\rm VCR_2}$ is the additional 
 variable combination $\{A,C\}$.}
We will come back to effective VCRs, when analyzing the incremental energy expressions for 
 calculating total energies of a molecular system from fragments and fragment combinations 
 (see Section \ref{sec:incr_en}).

\subsection{Uncoupled variables \label{sec:uncoupled_var}}
In the derivation of the double incremental scheme with semi-local coordinates, 
 we will meet cases, where variables of a function are uncoupled. 
In the following, we will give an induction argument, why the \chan{bar functions}, containing uncoupled 
 variables are zero.
We denote the uncoupled variable $y$ and let $l$ be the number of other variables from which 
 $y$ is uncoupled. Throughout this section, we assume the VCR to be closed on forming subsets.

Considering first the $l=1$ case, we denote  the two variables by $x$ and $y$ and
 call them  uncoupled in F if 
\begin{align}
F(x,y) = F(x,0) + F(0,y) - F(0,0).
\label{eq:uncoupled_pair}
\end{align}
The corresponding bar functions can be expanded according to Eqs.~(\ref{eq:incr_bar_combi}) and 
 (\ref{eq:incr_bar_from_contr}). 
 We obtain 
 $\bar{F}_1(x) = F(x,0) - F(0,0)$ and $\bar{F}_2(y) = F(0,y) - F(0,0)$ for the 1D functions. 
 For the 2D function we find that
$\bar{F}(x,y) = F(x,y) - \bar{F}_1(x) - \bar{F}_2(y)- F^0  
= F(x,0) + F(0,y)  - F(0,0) - F(x,0) + F(0,0) - F(0,y) + F (0,0) - F(0,0) = 0.$
Thus, the two-dimensional bar function with two uncoupled variables is exactly zero. 

We assume the induction hypothesis $\bar{F}(\{x\}^{{\mbf c}_s},y) = 0$ 
 for $s=l-1$ with $s,l\in \mathbb{N}$. 
We then set out to show that this also implies that $\bar{F}(\{x\}^{{\mbf c}_l},y) = 0$.
Again, $y$ is uncoupled from a set $\{x\}^{{\mbf c}_l}$ of other coordinates, and we 
 can write,
 similar to Eq.~(\ref{eq:uncoupled_pair}), 
\begin{align} \label{multiuncoup}
{F}(\{x\}^{{\mbf c}_l}, y)
= F(\{x\}^{{\mbf c}_l},0) + F(\{0\}^{{\mbf c}_l},y) - F(\{0\}^{{\mbf c}_l},0).
\end{align}
\chan{Introducing Eq.~(\ref{multiuncoup}) into Eq.~(\ref{eq:incr_bar_combi}) and separating the VCR summation in three terms, we obtain}
\begin{align}
\bar{F}^{{\mbf c}_l, y}(\{x\}^{{\mbf c}_l}, y)
=
 &  F(\{x\}^{{\mbf c}_l},0) + F(\{0\}^{{\mbf c}_l},y)
- F(\{0\}^{{\mbf c}_l},0) \notag \\
&
- 
\sum_{{\mbf c}_s \subseteq {\mbf c}_l}
\bar{F}^{{\mbf c}_s}(\{x\}^{{\mbf c}_s})
- 
\sum_{{\mbf c}_s \subset {\mbf c}_l, {\mbf c}_s \ne \emptyset}
\bar{F}^{{\mbf c}_s,y}(\{x\}^{{\mbf c}_s},y)
- \bar{F}(y).
\end{align}
Exploiting that the value of $ F(\{x\}^{{\mbf c}_l},0)$ is equal to that of 
 $ F(\{x\}^{{\mbf c}_l} ) 
= \sum_{{\mbf c}_s \subseteq {\mbf c}_l}
 \bar{F}^{{\mbf c}_s}(\{x\}^{{\mbf c}_s})  $ for 
 a given $\{x\}^{{\mbf c}_l}$  and 
$ \bar{F}(y) =   F(\{0\}^{{\mbf c}_l},y)  - F(\{0\}^{{\mbf c}_l},0) $, one sees
  that the only contributions to the bar function of a variable combination with an uncoupled
 variable arise from the lower-level variable combinations with one uncoupled variable,
\begin{align}
\bar{F}^{{\mbf c}_l, y}(\{x\}^{{\mbf c}_l}, y)
=
- 
\sum_{{\mbf c}_s \subset {\mbf c}_l, {\mbf c}_s \ne \emptyset}
\bar{F}^{{\mbf c}_s,y}(\{x\}^{{\mbf c}_s},y),
\end{align}
which is zero by the induction assumption.
One can, hence, conclude by induction  that this also holds for any variable combination   
 \chan{that contains} a  variable
 that is uncoupled to the others.

\subsection{Application of the incremental expansion to potential energy surfaces and evaluation of electronic energies}

\subsubsection{Incremental expansion of the potential energy surface}
The application of this framework to multi-dimensional PESs is equivalent to 
  the well-known $n$-mode expansions of the PESs outlined in the 
 Introduction.
The PES 
 is expressed in our notation (similar to that in Ref.~\onlinecite{kong2006}) as,
\begin{equation}
V(\{q\}) 
\approx 
\sum_{{\bf m}_n \in \{{\rm MCR}\}}  \bar{V}^{{\mbf m}_n} (\{q\}^{{\mbf m}_n})
=
\sum_{{\bf m}_n \in \{{\rm MCR}\}}  
\sum_{\chan{\substack{{{\mbf m}_{n'} \subseteq {\mbf m}_{n}}\\ {\mbf m}_{n'} \in \{{\rm MCR\}}}}}
k^{{\mbf m}_{n'},{\mbf m}_n} 
{V}^{{\mbf m}_{n'}} (\{q\}^{{\mbf m}_{n'}})
.
\label{eq:n-mode-rep}
\end{equation}
The coordinates $\{q\}$ can describe all possible arrangements of the nuclei. 
They may be in principle any set of such non-redundant coordinates.
${\bf m}_n$ is a composite index labeling a mode combination, which is 
 a set of $n$ modes.
All mode combinations that are explicitly parameterized in a given expansion of 
 \chan{$V(\{q\})$}
 are then collected in the so-called  mode combination range (MCR), which
 is accordingly the set of mode combinations.
$\{q\}^{{\mbf m}_n}$ is the subset of all 
 $n$
 vibrational coordinates in $\{q\}$ that are contained in the mode combination ${\mbf m}_n$.
The energy origin can be chosen freely, we assume $V^0=0$ at the expansion 
 point. 
\chan{Accordingly, the applied nomenclature is in complete analogy to the more general one 
 introduced above.}

\subsubsection{Incremental expansion of the electronic energy in fragmented approaches \label{sec:incr_en}}

The electronic energy of a system consisting of fragments can be expressed in an incremental 
 expansion in terms of its fragments. 
This has often been denoted by many-body expansion.
We can formulate this expansion of the electronic energy  
  similar to the one derived for PESs in Section II D 1:  
Assuming a system built up from $N$ fragments, we define
 $E=E(z_1,z_2,\dots,z_N)$ as the energy of the total system.
$z_1$, $z_2$, and $z_N$ are composite indices representing 
  particular conformations of the fragments $1$, $2$, and $N$, respectively.
The zero values of these composite indices $z_i = 0$ are defined such that 
 in this case no atoms of fragment $i$ are present in the respective conformation.
We further define the energy of a fragment combination in the conformation 
 $\{z\}_{{\mbf f}_l}$ as $E_{{\mbf f}_l}(\{z\}_{{\mbf f}_l})$ 
 and set the expansion point to $E_0=0$.  
\chan{Again, we employ the above-introduced nomenclature, but now the variables 
 describe the conformation of the individual fragments.
This means we expand the total energy in the framework described above  
 by setting up a \emph{fragment combination range} (FCR) containing all considered 
 fragment combinations ${\mbf f}_l$.}
It reads [analogously to Eq.~(\ref{eq:general_incr_expansion})]
\begin{equation}
 E(z_1,z_2,\dots,z_N) \approx 
 \sum_{{\mbf f}_l \in \{{\rm FCR} \}} \ubar{E}_{{\mbf f}_l} (\{z\}_{{\mbf f}_l}),
\label{eq:incr_energy}
\end{equation}
where 
\begin{align}
\ubar{E}_{{\mbf f}_l}  (\{z\}_{{\mbf f}_l})
  & =  {E}_{{\mbf f}_l} (\{z\}_{{\mbf f}_l}) - \sum_{\substack{{\mbf f}_{l'} \subset {\mbf f}_l \\  {\mbf f}_{l'} \in \chan{\{{\rm FCR}\}}}} 
 \ubar{E}_{{\mbf f}_{l'}} (\{z\}_{{\mbf f}_{l'}})
\label{eq:energy_bar}
\end{align}
 is the contribution of the fragment combination ${\mbf f}_l$ to the overall energy and
 is evaluated analogously to Eq.~(\ref{eq:incr_bar_combi}).
This expression 
 is equivalent to the incremental formulations being very widespread in quantum 
 chemistry, as outlined in the Introduction.
For the fragment combinations we use  subscripts instead of superscripts 
 for consistency to Sections II E -- II G.  
Note, that we also use underbars instead of bars to indicate corrections with respect 
 to lower-order fragment combinations.

Eqs.~(\ref{eq:incr_energy}) and (\ref{eq:energy_bar}) represent a
 very general energy expression for incremental molecular fragmentation approaches:
The energy expressions for different standard approaches are obtained for different 
 choices of FCR.
This formulation is thereby related 
 to other attempts to generalize the many-body expansion such as the generalized many-body (GMB) expansion by Richard and Herbert\cite{rich2012b} 
 or the many-overlapping-body  (MOB) expansion by Mayhall and Raghavachari\cite{mayh2012}.
The starting points of the different approaches are rather 
 different in the sense that both the GMB and MOB 
 expansions consider overlapping fragments, whereas we always have disjoint fragments
 as smallest units.
These are combined to larger, overlapping fragment combinations 
 in the incremental expansion. 
A detailed comparison of these formulations  
 and full account of the different fragmentation schemes covered 
 clearly goes beyond the scope of the present article. 
Still, we would like to reason that our scheme, 
 which starts from disjoint fragments, 
 can describe standard fragmentation methods with overlapping fragments as well.
Here, we use similar arguments as discussed in Ref.~\onlinecite{coll2015}.
The energy expression considering overlapping fragments often contains contributions 
 for overlaps of fragments. 
Assume we have a system ABC, to which we assign overlapping fragments AB and BC. 
A typical approach in fragmentation schemes with overlapping fragments would then 
 be to add the energies of AB and BC and 
 subtract the energy of the overlap, i.e., that of B.
The energy for this example is, hence, calculated as
  $E \approx E_{AB} +  E_{BC} - E_{AB}\cap E_{BC} = 
 E_{AB} +  E_{BC} - E_B $.
In our framework, we start out from non-overlapping fragments, 
 i.e., A, B, and C and then set up 
 a FCR, which is closed on forming subsets. 
Such an FCR could, for example read 
 ${\rm FCR} =\{\{ \}, \{A\}, \{B\}, \{C\}, \{A,B\}, \{B,C\} \}$.
Employing Eq.~(\ref{eq:coef}), we can determine the coefficients of all contributions 
 of the individual terms, i.e., of the cut functions.
This reveals, that the only non-zero contributions in this case 
 are $p_{\{A,B\}}^{\rm FCR} = p_{\{B,C\}}^{\rm FCR} = 1$ and $p_{\{B\}}^{\rm FCR} = -1$, so that the overall 
 energy expression is $E \approx E_{AB} +  E_{BC} - E_B $  and thereby equivalent to that
 obtained with overlapping fragments.
We anticipate, that the translation of the fragmentation approaches with overlapping 
 fragments to that presented here can generally be obtained by (i) defining disjoint 
 fragments from the \chan{overlaps of the fragments}, (ii) identifying 
 the largest fragment combinations to be considered in the expansion, 
 (iii) setting up a FCR that is closed on forming subsets, and 
 (iv) calculating the coefficients for each fragment combination 
 according to Eq.~(\ref{eq:coef})
 and thereby reduce the FCR, that is closed on forming subsets, to an effective FCR.

\chan{In this sense, incremental fragmentation approaches using overlapping fragments} 
 correspond to the incremental energy expansion used here with spatial 
 constraints to the considered coupling terms. 
In our terminology, this corresponds to a truncation of the FCR with 
 spatial considerations.
If we, for example, consider a chain like system ABCD, the FCR 
  up to three-fragment interactions reads
\begin{align} 
{\rm FCR^{3F}}
=
\{\{\},\{A\},\{B\},\{C\},\{D\},\{A,B\},\{A,C\},\{A,D\},\{B,C\},\{B,D\},\{C,D\}, \notag \\ \{A,B,C\},\{A,B,D\},\{A,C,D\},\{B,C,D\} \}.
\end{align} 
Restricting this FCR to direct neighbor coupling, we obtain
\begin{align} 
{\rm FCR^{3F,NB}}
=
\{\{\}, \{A\},\{B\},\{C\},\{D\},\{A,B\},\{A,C\},\{B,C\},\{B,D\},\{C,D\}, \notag \\ \{A,B,C\},\{B,C,D\} \},
\label{eq:FCR_spatialexample}
\end{align}
where $\{A,C\}$ and $\{B,D\}$ are not direct neighbors,
 but have to be included to ensure 
 that the FCR is closed on forming subsets. 
Such spatial restrictions of the FCR will be essential to obtain 
 linear scaling approaches, since each fragment has a limited number 
 of neighbors.
In many cases of these spatially restricted FCRs, it will be possible to 
 omit certain lower-order fragment combinations in the expansion based on Eq.~(\ref{eq:coef}), 
 see also Appendix~\ref{app:chain-eff}.
Computational savings due to the exploitation of such effective FCRs, 
 will be comparably small, 
since 
 such considerations only allow to neglect lower-order 
 fragment combinations and, thus, do not affect the 
 leading terms in the computational cost. 
The latter arise from the highest-order fragment combinations in the FCR.
See also the more detailed discussion on the scaling behavior of the different double 
 incremental schemes for generating PESs in Section~\ref{sec:scaling}.

\subsubsection{Translation of a point on the potential energy surface to the energy of a 
 given conformation}
So far, we have expressed the PES as well as the 
 individual energy points in incremental expansions. 
\chan{ Note, that we have employed two different sets of coordinates.
These are $\{z\}$, defining the actual 
 conformation, and $\{q\}$, a set of generalized internal vibrational coordinates.
The latter 
 specifies the displacement from a reference structure.}
Since we can express both by the other, i.e., write ${\mbf z}={\mbf z}({\mbf q})$ 
 and ${\mbf q}={\mbf q}({\mbf z})$, these 
 are interchangeable representations.
In case of rectilinear coordinates, we can express the positions (${\mbf z}_i$) of all $n$ nuclei
 in terms of nuclear positions for a reference configuration (${\mbf r}_0$)
 and rectilinear displacements (${\mbf d}_i$) as ${\mbf z}_i = {\mbf r}_0+ {\mbf d}_i$. 
These displacements can then be collected in a displacement vector 
 ${\mbf d}^T = (d_{1x}, d_{1y}, d_{1z}, d_{2x}, d_{2y}, d_{2z}, \dots , d_{nx}, d_{ny}, d_{nz})$.
Rectilinear vibrational coordinates can then be obtained by orthogonal  transformation 
 of the mass-scaled Cartesian displacement vectors\chan{ (${\mbf d}_{\rm m} = {\mbf M}^{\frac{1}{2}} {\mbf d}$) by  
 ${\mbf q} = {\mbf L}^T{\mbf d}_{\rm m} $}.
 ${\mbf M}$ is 
 a diagonal matrix of the size $3n\times 3n$ set up from $n$ diagonal $3\times 3$ matrices of
 the type $m_i {\mbf 1}$, where $m_i$ is the mass of the \chan{respective nucleus}.
With different orthogonal transformation matrices ${\mbf L}$, where ${\mbf L}^T{\mbf L}={\mbf 1}$,
 different sets of coordinates can be constructed. 
The column vectors of ${\mbf L}$, denoted by ${\mbf l}_k$, for $k=1,\dots,3n$, define 
 the corresponding coordinates as
\begin{align}   
q_k = {\mbf l}_k^T {\mbf d}_{\rm m}.
\label{eq:coord_conn}
\end{align}  
Accordingly, a displacement along vibrational coordinates ${\mbf q}$ 
 is directly related to a certain configuration of 
 the system by
\begin{align} 
{\mbf z} = {\mbf r}_0 + {\mbf M}^{-\frac{1}{2}}{\mbf L}{\mbf q},
\label{eq:coords-modes}
\end{align} 
defining a clear transformation between these variables.
In the discussion of the FALCON coordinates and PES transformations below, we will 
 use the ${\mbf L}$ matrices and their column vectors ${\mbf l}_k$ to describe coordinates, keeping 
 in mind that Eq.~(\ref{eq:coord_conn}) provides the rigorous connection.

\subsection{Double incremental expansion of the potential energy surface}
In the following, 
 the above incremental expansions of the PES and the 
 energy will be combined in a general manner, i.e., for any  MCR and FCR 
 satisfying the condition to be closed on forming subsets.
To establish a 
 connection between the PES points and the energy evaluations,
 we first define the value of a sub PES for the mode combination 
 ${\mbf m}_{n}$  at a grid point $\{q\}^{{\mbf m}_{n}}$ \chan{by the difference of its 
 electronic energy 
 to the energy of the reference structure $E({\mbf r}_0)$},
\begin{equation} \label{vtoedelta}
V^{{\mbf m}_{n}} (\{q\}^{{\mbf m}_{n}})
= E (\{z([q]^{{\mbf m}_{n}})\}) -  E({\mbf r}_0) 
= \Delta E (\{z([q]^{{\mbf m}_{n}})\}),
\end{equation}
where $[q]^{{\mbf m}_{n}} = \{\{q\}^{{\mbf m}_{n}},\{0\}^{{m \notin } {{\mbf m}_{n}}}\}$
 is a shorthand notation for a full set of modes where only those in ${\mbf m}_{n}$ can be 
 different from zero.

We then 
 insert an incremental expansion for the energy of the displaced as well as reference 
 structure, as given in Eq.~(\ref{eq:incr_energy}), employing the same FCR for both incremental 
 energy evaluations,
\begin{equation}
{V}^{{\mbf m}_n}(\{q\}^{{\mbf m}_n}) \approx 
\sum_{{\mbf f}_l \in \{{\rm FCR }\}} 
 \left[
\ubar{E}_{{\mbf f}_{l}} (\{{z}([q]^{{\mbf m}_{n}})\}_{{\mbf f}_l}) - \ubar{E}_{{\mbf f}_{l}} ({\mbf r}_{0,{\mbf f}_{l}})
 \right]
= \sum_{{\mbf f}_l \in \{{\rm  FCR }\}} 
\Delta \ubar{E}_{{\mbf f}_{l}} (\{z ([q]^{{\mbf m}_{n}})\}_{{\mbf f}_l})
\label{eq:subpot_incr}
\end{equation}
where ${\mbf f}_l$ is a fragment combination \chan{of} size $l$ in the 
 FCR and the sum runs over all fragment combinations 
 in the FCR. 
\chan{${\mbf r}_{0,{\mbf f}_{l}}$ denotes the reference conformation for the 
 fragment combination ${\mbf f}_{l}$.}
The corrected energy for a given fragment combination 
 can then be expressed by using Eq.~(\ref{eq:incr_bar_from_contr}) as 
\begin{equation}
\Delta \ubar{E}_{{\mbf f}_{l}} (\{z ([q]^{{\mbf m}_{n}})\}_{{\mbf f}_l})
=
\sum_{\substack{
{\mbf f}_{l'} \subseteq {\mbf f}_{l}\\
{\mbf f}_{l'} \in \{{\rm FCR}\}
}}
k_{{\mbf f}_{l'},{\mbf f}_{l}}
\Delta E_{{\mbf f}_{l'}} (\{z [q]^{{\mbf m}_{n}})\}_{{\mbf f}_{l'}}).
\label{eq:incr_energy_diff}
\end{equation}

We can then combine the above-obtained expressions  
 Eq.~(\ref{eq:incr_energy_diff}), Eq.~(\ref{eq:subpot_incr}), and Eq.~(\ref{eq:n-mode-rep}) \chan{to}
 the double incremental expression for the PES,
\begin{align}
V(\{ q\})
 & \approx
\sum_{{\mbf m}_n \in \{{\rm MCR} \}}
\sum_{\substack{
{\mbf m}_{n'} \subseteq {\mbf m}_{n}\\
{\mbf m}_{n'} \in \{{\rm MCR}\}
}}
k^{{\mbf m}_{n'}, {\mbf m}_{n}}
\sum_{{\mbf f}_l \in \{{\rm FCR} \}}
\sum_{\substack{
{\mbf f}_{l'} \subseteq {\mbf f}_{l}\\
{\mbf f}_{l'} \in \{{\rm FCR}\}
}}
k_{{\mbf f}_{l'}, {\mbf f}_{l}}
\Delta E_{{\mbf f}_{l'}} (\{z ([q]^{{\mbf m}_{n'}})\}_{{\mbf f}_{l'}}).
\label{eq:double_incr_1}
\end{align}
For later usage, we introduce the shorthand notation
\begin{align}
\Delta E^{{\mbf m}_{n'}}_{{\mbf f}_{l'}} 
\equiv
\Delta E_{{\mbf f}_{l'}} (\{z ([q]^{{\mbf m}_{n'}})\}_{{\mbf f}_{l'}}).
\label{eq:short}
\end{align}
With this, we write the general double incremental expansion of the PES as,
\begin{align}
V(\{ q\})
 & \approx
\sum_{{\mbf m}_n \in \{{\rm MCR} \}}
\sum_{\substack{
{\mbf m}_{n'} \subseteq {\mbf m}_{n}\\
{\mbf m}_{n'} \in \{{\rm MCR}\}
}}
k^{{\mbf m}_{n'}, {\mbf m}_{n}}
\sum_{{\mbf f}_l \in \{{\rm FCR} \}}
\sum_{\substack{
{\mbf f}_{l'} \subseteq {\mbf f}_{l}\\
{\mbf f}_{l'} \in \{{\rm FCR}\}
}}
k_{{\mbf f}_{l'}, {\mbf f}_{l}}
\Delta E^{{\mbf m}_{n'}}_{{\mbf f}_{l'}}.
\label{eq:double_incr_2}
\end{align}
The individual contributions to this expansion are, hence, energy differences for a 
 given mode combination (${\mbf m}_{n'}$) in a given fragment combination (${\mbf f}_{l'}$).
So far, we \chan{have interpreted} the double incremental expansion as an $n$-mode expansion with 
 incremental energy evaluations.
We will see in the following\chan{,} that breaking the \chan{PES contributions} down 
 to the basic terms of $\Delta E^{{\mbf m}_{n'}}_{{\mbf f}_{l'}}$ will give us the flexibility to reformulate the double incremental expansion 
 as an incremental expansion of the PES built up from PESs of contributing 
 fragment combinations.
In the following, it will be shown, that this interpretation in combination with semi-local coordinates will enable 
 the generation of full PESs with beneficial scaling. 

\subsection{Semi-local coordinates \label{sec:semi-local-coord}}

To achieve beneficial computational scaling with the double incremental scheme, 
 we need to use semi-local coordinates as outlined below. 
The defining characteristic of such coordinate sets is that some of the coordinates are constrained to
 internal motions in 
 a limited set of atoms and are thereby only \emph{active} for a limited number of fragments and 
 \emph{rigid} in others. 
The term rigid is used to underline that the \emph{internal} structure within these fragments 
 stays unchanged by the displacement, but the \emph{relative orientation} to other 
 fragments may be varied.
The discussion below on the usage of semi-local coordinates 
 in the double incremental expansion of PESs
 holds for all types 
 of coordinates fulfilling these criteria, regardless of the exact details on how these coordinates 
 are set up.

\begin{figure}
\includegraphics[width=\textwidth]{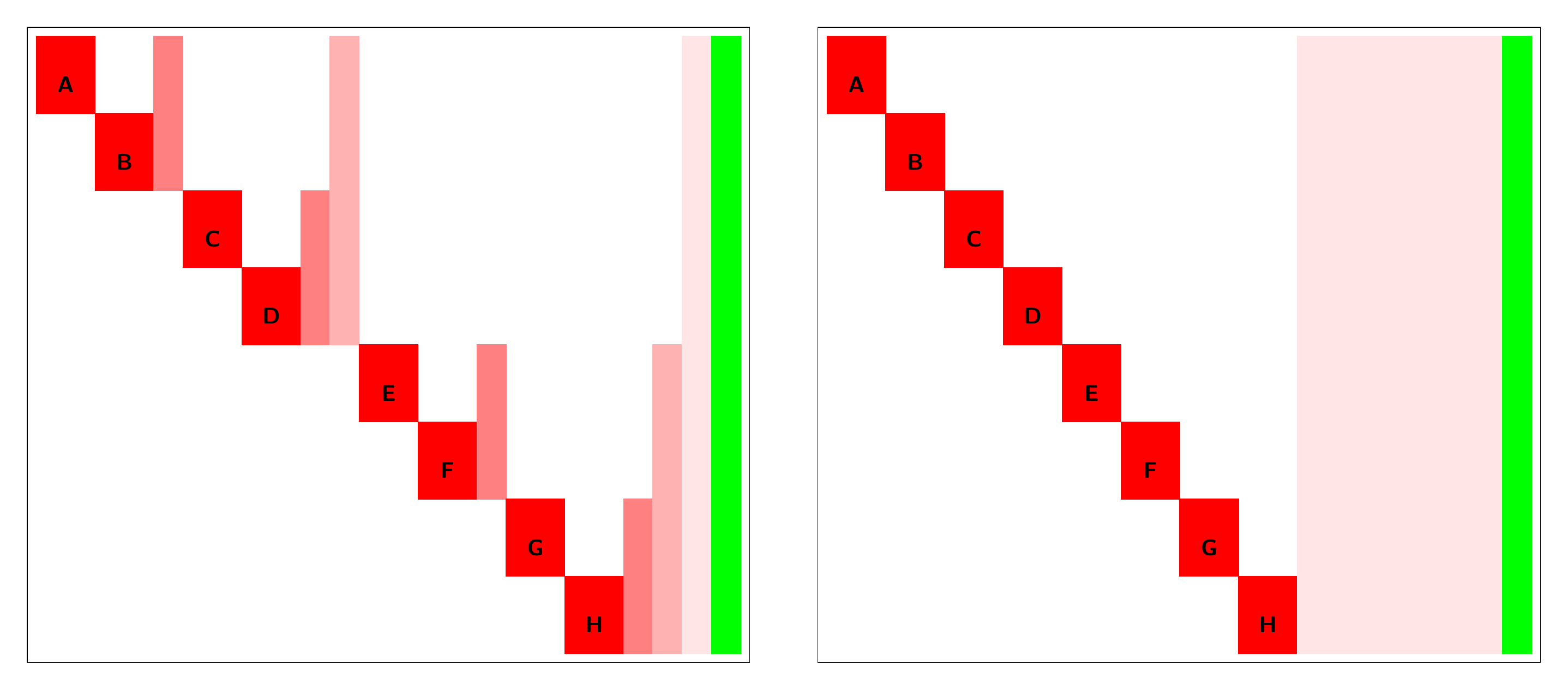}
\caption{\chan{
Spatial character of FALCON coordinates for two examples with eight fragments (A, B, C, D, E, F, G, and H)
with semi-local (left) and delocalized (right) inter-connecting modes.}
The squares symbolize the ${\mbf L}$ matrices for these FALCON coordinates.
These matrices only have non-zero entries on the colored areas.
The red blocks symbolize semi-local vibrational coordinates, where the deepest red
 indicates the largest degree of localization and the green areas
 correspond to the overall translational and rotational coordinates.
\label{fig:falcon_matrices}
}
\end{figure}

The set of semi-local coordinates used in this study are the recently introduced FALCON 
 coordinates\cite{koni2016}.
The FALCON algorithm enables the generation of  a full, rectilinear 
 set of purely vibrational coordinates with well-defined spatial character.
Possible structures of the ${\mbf L}$ matrix for such FALCON coordinates are 
 sketched in Figure~\ref{fig:falcon_matrices}.
Some of the FALCON coordinates only have contributions from a well-defined
 group of atoms, i.e., they are localized to a certain fragment\chan{
 (indicated by the deepest red in Figure~\ref{fig:falcon_matrices}}).
All other,\chan{ so-called} \emph{inter-connecting} coordinates span more than
 one fragment and are depicted in lighter red in Figure~\ref{fig:falcon_matrices}.
Depending on the settings of the FALCON scheme, 
 \chan{the inter-connecting 
 coordinates might} all span the entire molecule or only 
 a few fragments.
The total number of these inter-connecting coordinates is, however, in all 
 cases smaller than or equal to $(F-1)6$, where $F$ is the total number of 
 fragments.
Their number is, thus, significantly smaller than the total number of modes.

The inter-connecting coordinates are special kinds of coordinates in the sense that 
 the fragments or groups of fragments move relative to each other as \emph{semi-rigid} groups.
We choose the term semi-rigid, since the inter-connecting coordinates arise from the local translational 
 and rotational degrees of freedom of the groups, and are hence rigid in the 
 infinitesimal description. 
In a rectilinear space, as applied in the FALCON scheme, however, the relative rotation of different groups 
 cannot be described
 as rotations of internally fully rigid groups for larger displacements. 
This is why these types of coordinates are not strictly rigid. 

We denote the inter-connecting coordinates that move the semi-rigid entities $A$ and $B$ relative to  
 each other by $(A)\leftrightarrow(B)$ and those that move the fragment pairs $(AB)$ and $(CD)$
 as semi-rigid entities by  $(AB)\leftrightarrow(CD)$. 
\chan{
Occasionally, we also need to account for the truly rigid fragments in the formulation.
We mark these fragments and fragment combinations by square brackets.}
Thereby we can differentiate between the inter-connecting modes for the 
  left and right examples, respectively, in Figure~\ref{fig:falcon_matrices}.
In the left case, the inter-connecting modes, i.e.,
(A)$\leftrightarrow$(B)-[CDEFGH],
 [AB]-(C)$\leftrightarrow$(D)-[EFGH],
 [ABCD]-(E)$\leftrightarrow$(F)-[GH],
 [ABCDEF]-(G)$\leftrightarrow$(H),
 (AB)$\leftrightarrow$(CD)-[EFGH],
 [ABCD]-(EF)$\leftrightarrow$(GH),
 and 
 (ABCD)$\leftrightarrow$(EFGH),
have distinct and different spatial structures. 
All inter-connecting modes in the right 
 case are of type (A)$\leftrightarrow$(B)$\leftrightarrow$(C)$\leftrightarrow$(D)$\leftrightarrow$(E)$\leftrightarrow$(F)$\leftrightarrow$(G)$\leftrightarrow$(H) and thereby span the entire system.

\chan{All considered FALCON} coordinates diagonalize a reduced mass-weighted Hessian matrix 
 for a subspace of the vibrational space of the entire system.
By this procedure, we can assign \emph{quasi-harmonic} frequencies to these FALCON coordinates
 calculated from the diagonal elements of the reduced problem in the usual manner.  
For a more detailed description of the character of FALCON coordinates, we refer to the 
 section II B in Ref.~\onlinecite{koni2016}.

\subsection{Double incremental expansion in semi-local coordinates \label{sec:di-semi-local}}

\chan{
The double incremental PES
  expansion, Eq.~(\ref{eq:double_incr_2}),} has been derived and can be interpreted as \chan{an} $n$-mode expansion
 with incremental energy evaluation.
Turning the order of summation in the double incremental expansion  around,
\begin{align}
V(\{ q\})
 & \approx
\sum_{{\mbf f}_l \in \{{\rm FCR} \}}
\sum_{\substack{
{\mbf f}_{l'} \subseteq {\mbf f}_{l}\\
{\mbf f}_{l'} \in \{{\rm FCR}\}
}}
k_{{\mbf f}_{l'}, {\mbf f}_{l}}
\sum_{{\mbf m}_n \in \{{\rm MCR} \}}
\sum_{\substack{
{\mbf m}_{n'} \subseteq {\mbf m}_{n}\\
{\mbf m}_{n'} \in \{{\rm MCR}\}
}}
k^{{\mbf m}_{n'}, {\mbf m}_{n}}
\Delta E^{{\mbf m}_{n'}}_{{\mbf f}_{l'}},
\label{eq:fullpot_fc_mc_1}
\end{align}
we see that we can equally well interpret this expansion as an incremental expansion of 
 PESs of molecular fragments. 
\chan{Therefore, we introduce 
 the PES for the fragment combination ${{\mbf f}_l}$, 
 which can be expanded in an $n$-mode expansion as
\begin{align}
{V}_{{\mbf f}_{l'}} (\{ q\}) &  \approx 
\sum_{
{\mbf m}_n \in \{{\rm MCR} \}
}
\sum_{\substack{
{\mbf m}_{n'} \subseteq {\mbf m}_{n}\\
{\mbf m}_{n'} \in \{{\rm MCR}\}
}}
k^{{\bf m}_{n'}, {\bf m}_{n}}
\Delta E^{{\mbf m}_{n'}}_{{\mbf f}_{l'}} 
 = 
\sum_{
{\mbf m}_n \in \{{\rm MCR} \}
}
\bar{V}^{{\mbf m}_n}_{{\mbf f}_l'} (\{q\}^{{\mbf m}_{n}})
,
\end{align}
}where $\bar{V}^{{\mbf m}_n}_{{\mbf f}_l} (\{q\}_{{\mbf m}_{n}})$
 is the corresponding bar potential for the mode combination ${\mbf m}_{n}$.
In this interpretation, it is striking that the PES 
 of every fragment combination 
 is spanned by the same number of modes as the total system.
This number is significantly higher than the number of degrees of freedom in the 
 respective fragment combination. 

Applying semi-local coordinates, however, all coordinates that are rigid within the 
 respective fragment combination are uncoupled from the other modes.
These have, hence,  
 zero \chan{contribution} to the respective bar potentials 
 $\bar{V}^{{\mbf m}_n}_{{\mbf f}_l} (\{q\}^{{\mbf m}_{n}})$, see also Section~\ref{sec:uncoupled_var}.
\chan{We
define ${\mbf m}_{r, {{\mbf f}_l}}$ as a set of modes 
 containing all $r$ modes that are rigid 
 in the fragment combination ${{\mbf f}_l}$, 
 i.e., do not change the internal structure of ${{\mbf f}_l}$.
With this,
we
 can restrict} the MCR for the PES of the fragment combination ${{\mbf f}_l}$,
\begin{align}
{V}_{{\mbf f}_l} (\{q\}) 
 = 
\tilde{V}_{{\mbf f}_l} (\{q\} \setminus \{q\}^{{\mbf m}_{r,{{\bf f}_l}}}) 
 \approx 
\sum_{
\substack{
{\mbf m}_n \in \{{\rm MCR} \} \\
{\mbf m}_n \cap {\mbf m}_{r,{{\mbf f}_l}} = \emptyset
}
}
\bar{V}^{{\mbf m}_n}_{{\mbf f}_l}
(\{q \}^{{\mbf m}_n})
,
\label{eq:subpot}
\end{align}
where the MCR has to be closed on forming subsets.
Here, the tilde on the V is simply to mark that this function is formally 
 different from the standard V functions since it is a function of fewer coordinates, but with the same values. 
Notice that in Eq.~(\ref{eq:subpot}) only modes contribute that are 
 non-rigid in the respective fragment combination.
We can now rewrite the full PES, 
 Eq.~(\ref{eq:fullpot_fc_mc_1}), as
\begin{align}
V(\{ q\})
& \approx
\sum_{{\mbf f}_l \in \{{\rm FCR} \}}
\sum_{\substack{
{\mbf f}_{l'} \subseteq {\mbf f}_{l}\\
{\mbf f}_{l'} \in \{{\rm FCR}\}
}}
k_{{\mbf f}_{l'},{\mbf f}_{l}}
\tilde{V}_{{\mbf f}_{l'}}
(\{q \} \setminus \{q\}^{{\mbf m}_{r,{\bf f}_{l'}}})  
=
\sum_{{\mbf f}_l \in \{{\rm FCR} \}}
\ubar{\tilde{V}}_{{\mbf f}_{l}}
(\{q \} \setminus \{q\}^{{\mbf m}_{r,{\bf f}_l}})  
\label{eq:fullpot_fc_mc_truncf}
,
\end{align}
This expansion of the PES is particularly appealing since it significantly 
 reduces the number of modes to be considered in the PESs for the 
 individual fragment combinations, when dealing 
 with semi-local modes. 
Thereby the computational cost of generating the PESs can be significantly   
 reduced as outlined in detail in Section \ref{sec:scaling}. 

\subsection{Concrete computational methods and computational scaling \label{sec:scaling}}
The computational cost of a PES generation is governed by the accumulated 
 computational costs of the required SPCs, which is 
determined by  the number of individual SPCs needed and the cost per SPC. 
Unfortunately, as the system size increases, both of these factors increase drastically.
 We will see that 
 the double incremental expansion proposed in this work can reduce the computational cost of 
 PES \chan{generations} and \chan{the} associated scaling with increasing system size.
For a static grid, we can directly calculate the number of required SPCs for the 
 different approaches. 
We assume that all of the fragments are non-linear and therefore have 
 six nuclear degrees of freedom from translational and rotational motions. 
We further assume throughout this section a chain-like system and account only for 
 direct neighbor couplings.
The assumed FALCON coordinates are of the same type as those shown in the right 
 part of Figure~\ref{fig:falcon_matrices}. 
In this way, we obtain $(F-1)6$ completely delocalized inter-connecting coordinates, where 
 $F$ is the total number of fragments.
We furthermore exploit the reduction of required terms by forming an
  effective FCR for \chan{these kinds of systems} 
 (see Appendix \ref{app:chain-eff}).
In the present case, only the largest fragment combinations contribute as well as the 
 second largest ones that do not contain an outer-most fragment \chan{of the chain-like system}.

\chan{In this section}, we will 
 develop the number of required SPCs  and formal computational 
 scaling for the different expansions of the PES assuming the particular setup
 of a chain-like system described above. 
  \chan{This is a significant simplification and many real systems will be far from this simple. However, 
  this simplification is convenient for the clarity of the argument. 
It is expected that even if the simplest nearest neighbor description is
  not fully adequate for all systems, most systems will be such that a given fragment has only significant interaction
  with a limited set of other fragments. As long as this limited set of fragments does not increase with the size of the system
  the analysis is still relevant for a rough scaling estimate, though not intended to be a precise prediction of the actual computational cost.  
The derived benefits for the double incremental schemes 
 compared to the conventional models are, hence, expected to be 
 valid for a wide range of systems. }

In the following equations, 
$N_{\rm pfr}$ stands for the  number of atoms per fragment, which is 
 assumed to be the same for every fragment,
$n$ is the maximal number of modes per mode combination, and 
$f$ is the maximal number of fragments per fragment combination.  
We use $g$ for the number of grid points per mode and $s$ as 
 scaling exponent for the underlying electronic structure method.
This means, 
 the cost of the electronic structure calculations is assumed to be proportional to
 $(N_{\rm at})^s$ for a system with $N_{\rm at}$ atoms.
For a given fragment combination of $l$ fragments, the computational cost will thus be 
 proportional to $(l \cdot N_{\rm pfr})^s$. 
Assuming $N_{\rm pfr}$ is constant, the interesting aspect is that the computational 
 cost for the most demanding SPC in a double incremental PES generation
 increases as $f^s$ with increasing maximal fragment combination level $f$.

\subsubsection{Conventional $n$-mode expansion \label{sec:scal_conv}}
For the conventional $n$-mode expansion the number of SPCs is given \chan{by}
\begin{align}
N_{\rm SPC}^{n{\rm -mode}}
=
\sum_{m=0}^n 
\begin{pmatrix}
F ( 3 N_{\rm pfr}-6) + (F-1)6 \\
m
\end{pmatrix}
g^m 
,
\label{eq:scale_full}
\end{align}
where the total number of vibrational modes ($ M_{\rm tot} = 3F  N_{\rm pfr} - 6$) is calculated  
 as $F$ 
 times the number of modes per fragment ($3 N_{\rm pfr}-6$) plus the 
 \chan{remaining part of $(F-1)6$ modes}.
The \chan{number of mode combinations of  mode-combination level $m$} 
 is then obtained as the binomial coefficient
 of $\begin{pmatrix}  M_{\rm tot} \\ m \end{pmatrix}$. 
Each of these mode combinations requires  $g^m$ SPCs.
In the limit of a large number of fragments, the number of SPCs
  scales with $F^n$. 
And every individual SPC exhibits \chan{an} $F^s$ scaling.  
The combined scaling is given as $F^{n+s}$.

\subsubsection{Double incremental expansion in normal modes (DIN) \label{par:convincrE}}
\chan{
Using the same $n$-mode expansion as in Section \ref{sec:scal_conv} but with incremental energy 
 calculations, the evaluation of the energy difference for a particular conformation
 requires one SPC 
 for each fragment combination that has a non-zero contribution to the overall energy.}
This is the case 
 for all fragment combinations in normal modes,
 since a normal mode will generally alter the internal structure of every fragment
 in the system. 
For the given setup of a chain-like system with neighbor coupling and a maximal fragment combination level of \chan{$f$}, we 
  need to consider  $(F-f+1)$ fragment combinations of size $f$ as well as 
 $(F-f)$  fragment combinations of size $(f-1)$.
\chan{The contributions of all lower-order fragment combinations are exactly zero in this 
 setup as outlined in Appendix \ref{app:chain-eff}.}
The number $(F-f)$ is obtained from the total number of $(F-(f-1)+1)$ connected fragment combinations of size $(f-1)$ 
 minus \chan{2}.
By this, we account for the fact, that the two outer-most fragment combinations have zero-contribution (see 
 Appendix~\ref{app:chain-eff}).
We can then calculate the overall number of required SPCs as
\begin{align}
N_{\rm SPC}^{\rm DIN}
=
\left[
\sum_{m=0}^n
\begin{pmatrix}
F (3 N_{\rm pfr}-6) + (F-1)6 \\
m
\end{pmatrix}
g^m 
\right].
\left[
\sum_{l=f-1}^f (F-l+ (-1)^{f-l})
\right]
.
\label{eq:scale_din}
\end{align}
We use the squared brackets to highlight that the number of grid points in the overall PES (left bracket) 
 is independent \chan{of} the number of fragment combinations in the FCR (right bracket). 
The latter number can be interpreted as the number of SPCs per grid point of the full system. 
An alternative \chan{viewpoint on} these two terms in brackets would be that the left bracket contains the 
 number of grid points or SPCs required to generate the PES of one fragment combination.
This number is then multiplied with the number of fragment combinations. 
In any case, the overall scaling in \chan{the number of} SPCs for large $F$ is $F^{n+1}$. 
\chan{The gain here lies} in the fact that the cost per SPC is largely reduced\chan{ and is 
 independent of F}. 
Accordingly, the scaling of the accumulated computational cost of all SPCs is likewise  $F^{n+1}$.
This scaling behavior is not limited to the double incremental scheme in normal modes. 
In fact, it holds for all types of coordinates that are not strictly rigid in the sense 
 that all modes can alter the internal structure of all fragments and fragment combinations.

\subsubsection{\chan{Double incremental expansion in FALCON coordinates} (DIF) \label{par:dif}}
When using FALCON coordinates, we can restrict the number of modes in the PESs for the 
 individual fragment combinations to those that are non-rigid in the respective 
 fragment combination, as shown in Section~\ref{sec:di-semi-local}.
\chan{Thereby we can reduce the number of modes to be considered in 
 a PES construction for} a fragment combination of size $l$ to 
 $l (3N_{\rm pfr}-6) + (F-1)6$, where we assume that all $(F-1)6$ inter-connecting modes contribute. 
The overall number of required SPCs is then reduced to
\begin{align}
N_{\rm SPC}^{\rm DIF}
=
\sum_{l=f-1}^f (F-l+(-1)^{f-l})
\sum_{m=0}^n
\left[
\begin{pmatrix}
l (3N_{\rm pfr}-6) + (F-1)6 \\
m
\end{pmatrix}
g^m
\right] 
.
\label{eq:scale_dif}
\end{align}
Note, that we cannot turn these sums freely around since the number of SPCs required in the 
 PES generation of the respective fragment combination depends on the fragment combination.
Despite the drastic reduction of the number of required SPCs, we obtain the same 
 overall formal scaling of $F^{n+1}$ as in Section \ref{par:convincrE}.
This is due to the fact that the maximal number of modes in a fragment combination still generally scales linear 
 with $F$.  We shall see shortly that DIF should be expected to provide a 
 \chan{much-reduced} computational effort compared to a standard approach.
 Nevertheless, we proceed to consider reducing the formal \chan{computational} scaling as well. 

\subsubsection{\chan{Double incremental expansion in  FALCON coordinates} with auxiliary coordinate transformation (DIFACT) \label{sec:difact}}
In Section \ref{par:convincrE}, the PES of a fragment combination is calculated in the same 
 number of modes as the complete system. 
The number of relevant modes for a fragment combination has then been 
  drastically reduced by introducing semi-local 
 FALCON coordinate in Section \ref{par:dif}. 
We now go one step further and introduce auxiliary inter-connecting modes, spanning the vibrational space 
 of the inter-connecting modes within each fragment combination in a non-redundant manner. 
For a fragment combination consisting of $l$ fragments, we have, next to the 
 $l (3N_{\rm pfr}-6)$ intra-fragment modes, $(l-1)6$ vibrational degrees of freedom, which 
 we express as inter-connecting modes.
In rectilinear coordinates, we have to include additional three coordinates describing the overall 
 infinitesimal rotation of the fragment combination.
This is due to the limitations of rectilinear vibrational coordinates 
 with respect to the description of ``real'' rotations 
 for non-infinitesimal displacements.  
The overall number of required SPCs is then obtained as
\begin{align}
N_{\rm SPC}^{\rm DIFACT}
=
\sum_{l=f-1}^f (F-l+(-1)^{f-l})
\sum_{m=0}^n
\left[
\begin{pmatrix}
l (3N_{\rm pfr}-6) + (l-1)6 + 3 \\
m
\end{pmatrix}
g^m
\right] 
.
\label{eq:scale_difact}
\end{align}
In this case, the number of considered modes for a certain fragment combination is determined by 
 the size of this fragment combination and independent of $F$, and so is the number of SPCs
 required to generate the PES of the respective fragment combination.
The number of \chan{fragment combinations to be considered}, however, 
 still scales linear with $F$, and the \chan{total} number of required SPCs scales likewise.
Since also here the cost per individual SPC is independent of $F$, the 
 linear scaling is also obtained when considering the accumulated cost for all 
 \chan{SPCs that are required when constructing the PES in a the double incremental manner.
The introduction of specific
 auxiliary coordinates for the generation of the PESs for individual fragment combinations},
 has, however, the disadvantage, 
 that it requires subsequent transformation of the PESs to the common, overall 
 coordinates.
Such a transformation is only exact in case of a full PES representation. 
Since this is typically not feasible, we will generally introduce a transformation error when 
 going from DIF to DIFACT. 
The implemented transformation step 
  exhibits polynomial scaling with the number of modes 
 (see Section \ref{sec:transform} for further details). 
Its overall cost is, however, typically orders of magnitudes lower than the 
 explicit calculation of the corresponding SPCs, and the computational cost of 
 the transformation has therefore so far not been a bottleneck.

\subsubsection{Example scaling behavior}
\begin{figure}
\includegraphics[width=0.8\textwidth]{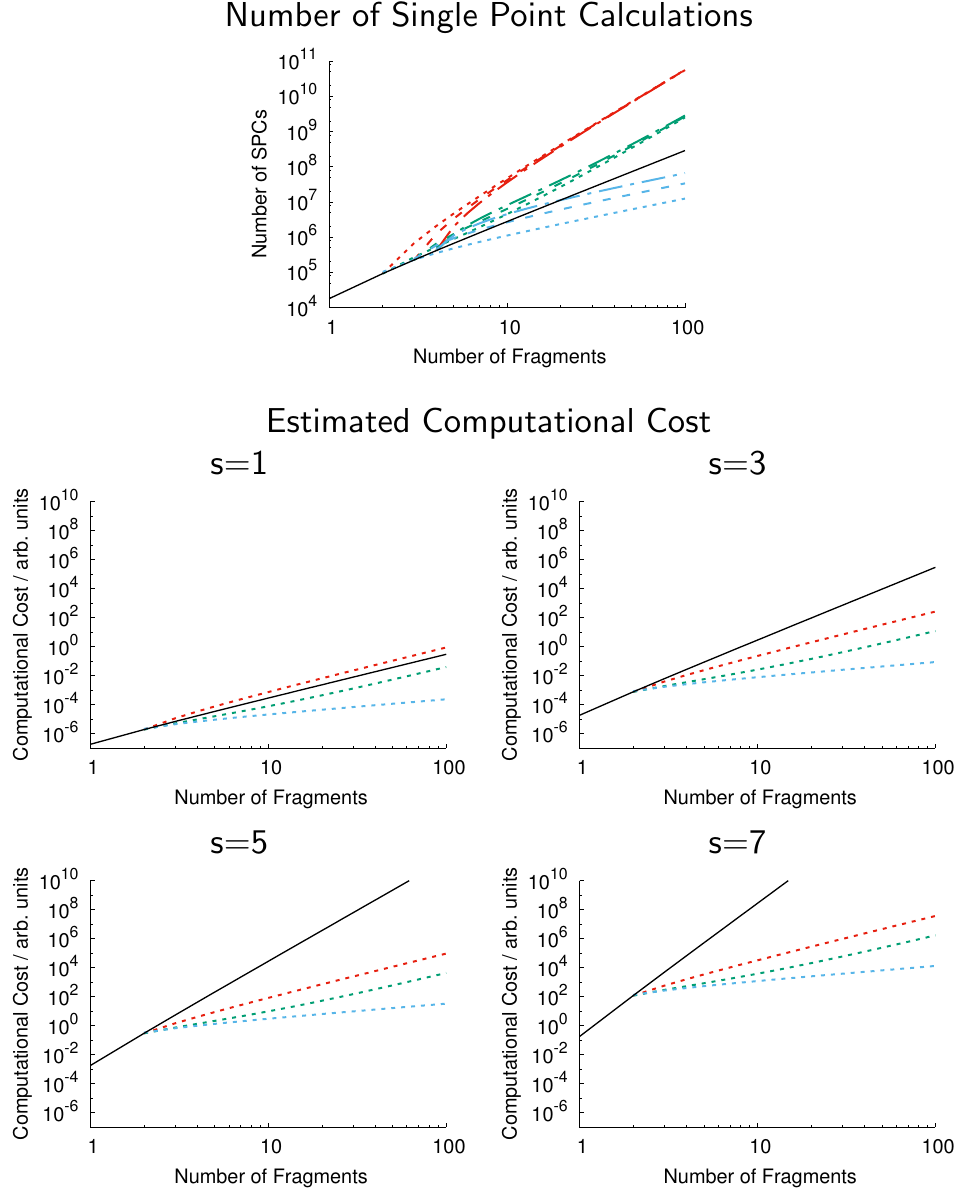}
\caption{
\chan{
Upper part:
Number of single point calculations (SPCs)
 for PES generations with the different schemes, i.e., conventional $n$-mode representation
 (black), DIN (red), DIF (green), and 
 DIFACT (blue) according to Eqs.~(\ref{eq:scale_full}),  (\ref{eq:scale_din}), (\ref{eq:scale_dif}),
 and (\ref{eq:scale_difact}), respectively, 
 for different number of fragments ($ F  =1-100 $) and different 
 fragment-combination level, i.e., $ f=2$ (- - -), $f=3$ ($- - $), and $f=4$ ($\cdot - \cdot$) 
 in the double incremental schemes
 distinguished by the line type.
The default values are
$ N_{\rm pfr}  = 10 $,
$ n  = 2           $,
$ g  = 8           $.
Bottom part:
Estimated computational cost obtained for $f=2$ by multiplying the 
 number of SPCs for a certain system size of $l\cdot N_{\rm pfr}$ atoms
 by $A\cdot (l \cdot N_{\rm pfr})^s$, where  $s=1,3,5,7$ and $A = 10^{-12}$.
It employs the same color code for the different methods as the upper part.
}
}
\label{fig:scaling}
\end{figure}

The number of SPCs for the different schemes with increasing number of fragments 
 is shown in Figure~\ref{fig:scaling}
 for a chain-like system assuming fragments of ten atoms (24 modes per fragment) 
 and eight grid points per mode.
In these examples, we assume the exact same type of system as in the 
 scaling considerations above.
It is seen that the variation in the actual number of SPCs matches well the 
 asymptotic predictions of $\log(\#SP) \approx \log(F)$ and $\log(\#SP) \approx (n+1)\log(F)$ 
 for DIFACT and DIN or DIF, respectively, where $n=2$ in the present example.
The number of required SPCs in the DIN and DIF schemes lies above the number 
 of SPCs required for the conventional treatment for all system sizes.
For the DIFACT scheme, this only holds for a small number of fragments.
In this scheme, the number of required SPCs increases 
 less drastically with the number of fragments 
 than for the conventional $n$-mode treatment.
The crossover points between the number of SPCs in DIFACT 
 with \chan{that for} the conventional $n$-mode expansion are at system sizes around four fragments for $f=2$
 and eleven fragments for  $f=3$.
 
The analysis so far is limited by the fact that the number of SPCs is not representative for the total computational cost, since the individual SPCs can have quite different cost and they may refer to 
 SPCs with different number of atoms. 
To estimate the computational cost for a PES generation, we have 
 \chan{weighted} 
 each SPC \chan{by} the factor $A \cdot (N_{\rm at, FC})^s$ mimicking the effect of the
 different computational cost.  $N_{\rm at, FC}$ is the
 number of atoms in the relevant fragment combination.
The resulting estimates for $s=1,3,5,7$ \chan{with $n=2$ and $f=2$} 
 are plotted in the four lower graphs in Figure~\ref{fig:scaling}.
These exponents have been chosen to represent the computational scaling of 
 electronic structure methods in standard implementations, i.e., third order for 
 density functional theory (DFT), fifth order for second-order M{\o}ller--Plesset (MP2)
 calculations and seventh order for coupled cluster with doubles and perturbative 
 triples [CCSD(T)]. 
Additionally, linear scaling approaches are considered.
Linear scaling methods have been a major focus of electronic structure development work
 in the last decades, and we have here included it to show clearly that even in this case
 the expected computational cost for the double incremental approach in FALCON coordinates 
 is lower than for the
 conventional $n$-mode expansion combined with a linear scaling algorithm for the SPCs. 
This also holds for the DIF scheme, which requires \chan{more SPCs}
 than the conventional $n$-mode expansion. 
The observation that
 the overall cost increases, when combining the DIN scheme with linear scaling approaches for the SPCs, 
 reflects that the application of the fragmentation approach to already linear scaling 
 electronic structure methods cannot reduce the scaling further, but introduces extra 
 costs.
This combination is not a typical application at any means. 

Besides a general increase in computational cost with increasing $s$, we also see that the 
 double incremental schemes get more and more beneficial for larger values of $s$.
For systems with more than a few fragments we find
 orders of magnitude efficiency gains in DIN, DIF, and DIFACT
 methods compared to the conventional $n$-mode expansion. 
\chan{Since} the individual \chan{SPCs} have similar cost in DIN,
 DIF, and DIFACT, the additional order of magnitude gains
 from DIN to DIF and DIFACT derives from the reduction in 
 the number of SPCs seen in the upper part of the figure.   
Not surprisingly, the additional computational cost, when increasing the
  maximal fragment combination level, 
 is larger the higher the scaling for the applied 
 electronic structure method is,  
 i.e., the larger $s$ \chan{(see Figure~S-1 in the supplementary material\cite{sm})}.

\subsection{Auxiliary coordinates}

As seen \chan{in Section \ref{sec:difact}}, we can obtain linear scaling of the accumulated cost 
 of the required SPCs in a PES generation, as long as the number of 
 modes to be accounted for in the generations of the PESs for the 
 individual fragment combinations
 does not scale with the overall number of fragments. 
Such a situation might be achievable in internal, curvilinear modes. 
In rectilinear coordinates, which allow easy evaluation of the kinetic energy 
 in vibrational structure calculations, we are not aware of 
 local purely vibrational coordinates that 
 fulfill this requirement generally. 
\chan{We propose} the usage of alternative auxiliary coordinates for the generation of 
 the PESs of the individual fragment combinations. 
These PESs in auxiliary coordinates need to be transformed to 
 the respective PES in the common coordinates 
 before the linear combination of the PES for the individual fragment combinations is possible.   
This transformation is in practice approximate (see Sections \ref{sec:transform} and 
 \ref{sec:app_transform}). 
The resulting transformation error can be expected to be smaller 
 the more the final coordinates are resembled by the auxiliary coordinates. 
This means that it is desirable to obtain the auxiliary coordinates in a similar manner as  
 the common inter-connecting coordinates.
In the present case, we use FALCON-type coordinates for the common 
 coordinates, which suggests 
 the use of a similar scheme for the generation of the auxiliary coordinates.
We, construct the auxiliary coordinates as FALCON-type inter-connecting coordinates 
 that are local to the respective fragment combination.
For a more detailed description of the applied protocol, we refer to Section~\ref{sec:prep}. 

\subsection{Transformation of  \chan{potential energy surfaces} of fragment combinations from auxiliary to common FALCON coordinates \label{sec:transform}}

With the PES transformation, we aim at a PES representation
 for a fragment combination (FC) in the local redundant coordinates that corresponds 
 to the common FALCON coordinates 
 (${\mbf L}^{\rm F}_{\rm FC}$).
The starting point of the transformation is a PES representation, in which 
 some of the vibrational degrees of freedom are expressed in 
 auxiliary coordinates ${\mbf L}^{\rm aux}_{\rm FC}$.
Let us first assume, our set of auxiliary coordinates includes vibrational
 inter-connecting (I), 
 as well as rotational (R) and translational (T) 
 coordinates of the fragment combination,
\begin{align}
{\mbf L}^{\rm aux}_{\rm FC} 
= \left( ^{\rm I}{\mbf L}_{\rm FC}^{\rm aux} | ^{\rm R}{\mbf L}_{\rm FC}^{\rm aux} | ^{\rm T}{\mbf L}_{\rm FC}^{\rm aux}\right).
\end{align}
We refer to Section~\ref{sec:prep} for a more detailed discussion on how 
 \chan{these coordinates} are obtained. 
 For now, we only need to know that the column vectors of ${\mbf L}^{\rm aux}_{\rm FC}$ are orthonormal. 
In this case, the complete space spanned by ${\mbf L}^{\rm F}_{\rm FC}$ is also contained in 
 ${\mbf L}^{\rm aux}_{\rm FC}$ and we can write
\begin{align}
{\mbf L}_{\rm FC}^{\rm F} = {\mbf L}^{\rm aux}_{\rm FC} {\mbf A}_{\rm FC},
\end{align}
where the rectangular transformation matrix is set up as
\begin{align}
{\mbf A}_{\rm FC} =  ({\mbf L}^{\rm aux}_{\rm FC})^T{\mbf L}_{\rm FC}^F
=
\begin{pmatrix}
^{\rm I}{\mbf A}_{\rm FC}  \\
^{\rm R}{\mbf A}_{\rm FC}  \\
^{\rm T}{\mbf A}_{\rm FC}  \\
\end{pmatrix}
, 
\end{align}
so that
\begin{align}
{\mbf L}_{\rm FC}^{\rm F} = 
{^{\rm I}}{\mbf L}_{\rm FC}^{\rm aux} \ {^{\rm I}}{\mbf A}_{\rm FC}
+ {^{\rm R}}{\mbf L}_{\rm FC}^{\rm aux}\  {^{\rm R}}{\mbf A}_{\rm FC}
+ {^{\rm T}}{\mbf L}_{\rm FC}^{\rm aux}\  {^{\rm T}}{\mbf A}_{\rm FC}
.
\label{eq:a_all_terms}
\end{align}
\chan{The molecular structure of a fragment combination for a
 displacement along these common coordinates around the reference point $(\mbf{r}_0)_{\rm FC}$ 
 can be obtained using Eq.~(\ref{eq:coords-modes})
 as 
}
\begin{align}
{\mbf z}_{FC}^{\rm F}
 = & (\mbf{r}_0)_{\rm FC} 
 + {\mbf M}^{-\frac{1}{2}}_{\rm FC} {\mbf L}^F_{\rm FC}{\mbf q}_{\rm FC}^{\rm F}
\notag
\\
 = & (\mbf{r}_0)_{\rm FC} 
 + {\mbf M}^{-\frac{1}{2}}_{\rm FC}\ {^I}{\mbf L}^{\rm aux}_{\rm FC}\ {^I}{\mbf A}_{\rm FC}{\mbf q}_{\rm FC}^{\rm F}
 + {\mbf M}^{-\frac{1}{2}}_{\rm FC}\ {^R}{\mbf L}^{\rm aux}_{\rm FC}\ {^R}{\mbf A}_{\rm FC}{\mbf q}_{\rm FC}^{\rm F}
 + {\mbf M}^{-\frac{1}{2}}_{\rm FC}\ {^T}{\mbf L}^{\rm aux}_{\rm FC}\ {^T}{\mbf A}_{\rm FC}{\mbf q}_{\rm FC}^{\rm F}
\notag
\\
 = & (\mbf{r}_0)_{\rm FC} 
 + \Delta{^{\rm I}}\mbf{r}_{\rm FC}^{\rm F}
 + \Delta{^{\rm R}}\mbf{r}_{\rm FC}^{\rm F}
 + \Delta{^{\rm T}}\mbf{r}_{\rm FC}^{\rm F}
  .
\end{align}
The subscript ${\rm FC}$ indicates that only the contributions of 
 the atoms in a particular fragment combination are considered, 
\chan{e.g., ${\mbf q}_{\rm FC}^{\rm F}$ denotes the part of the common FALCON coordinates 
 that effect the fragment combination FC.
 $\Delta{^{\rm I}}\mbf{r}_{\rm FC}^{\rm F}$,
 $\Delta{^{\rm R}}\mbf{r}_{\rm FC}^{\rm F}$, and 
 $\Delta{^{\rm T}}\mbf{r}_{\rm FC}^{\rm F}$
 are} the displacements within this fragment combination  
 along the common FALCON coordinates
 due to vibrational inter-connecting, rotational, and translational motions, 
 respectively. 
The energy of a fragment combination ${\rm FC}$ 
 for a particular displacement along ${\mbf L}_{\rm FC}^{\rm F}$, can, hence, be
 expressed as
\begin{align}
E_{\rm FC}(\{z(\{q [{\mbf L}_{\rm FC}^{\rm F}] \}) \}_{\rm FC})
& = 
E_{\rm FC}(
  (\mbf{r}_0)_{\rm FC}
 + \Delta{^{\rm I}}\mbf{r}_{\rm FC}^{\rm F}
 + \Delta{^{\rm R}}\mbf{r}_{\rm FC}^{\rm F}
 + \Delta{^{\rm T}}\mbf{r}_{\rm FC}^{\rm F}
 ) \notag \\
& = 
E_{\rm FC}(
  (\mbf{r}_0)_{\rm FC}
 + \Delta{^{\rm I}}\mbf{r}_{\rm FC}^{\rm F}
 + \Delta{^{\rm R}}\mbf{r}_{\rm FC}^{\rm F}
 )  
\notag
 \\
& = 
E_{\rm FC}(\{z(\{q [^{\rm I}{\mbf L}_{\rm FC}^{\rm aux}\  {^{\rm I}}{\mbf A}_{\rm FC}
+ {^{\rm R}}{\mbf L}_{\rm FC}^{\rm aux}\  {^{\rm R}}{\mbf A}_{\rm FC} ] \}) \}_{\rm FC}),
\label{eq:e_a_all_terms}
\end{align}
where $\{q[\mbf{L}]\}$ expresses that the set of coordinates $\{q\}$ is defined through the 
 $\mbf{L}$ matrix. 
The second equality in Eq.~(\ref{eq:e_a_all_terms}) holds, due to the translational invariance of the energy.
As previously noted, the rotational coordinates
will not generally be \chan{purely} rotational, and we cannot
do a similar trick for these as long as we are
  using rectilinear coordinates. 
Since all points on the PES are calculated from the 
 electronic energy points, 
 similar arguments can  be used to express
\begin{align}
 V_{\rm FC}(\{q [{\mbf L}_{\rm FC}^{\rm F}] \})
&  = 
{V}_{\rm FC}(\{q [^{\rm I}{\mbf L}_{\rm FC}^{\rm aux} \  {^{\rm I}}{\mbf A}_{\rm FC}
+ {^{\rm R}}{\mbf L}_{\rm FC}^{\rm aux}\  {^{\rm R}}{\mbf A}_{\rm FC} ]\}).
\end{align}
Thus,
 it is sufficient to include internal and rotational coordinates in the auxiliary 
 set of coordinates. 
By default we apply auxiliary coordinates and transformation matrices that do not 
 consider the translational contributions

The general transformation algorithm using the respective 
 transformation matrices can be found in Section \ref{sec:app_transform}
 for PESs represented in polynomials of the coordinates. 
We note that whenever we transform polynomial PES representations with higher polynomial 
 order than mode combination level, the transformation formally leads to 
 higher-order mode combinations that are not included in the original PES
 representation\cite{yagi2012}.
Similar arguments can also be applied the other way around: higher-order
 mode combinations in auxiliary coordinates
 could upon transformation contribute
 to lower-order mode combinations in the common FALCON coordinates.
All together, this means that the transformation of the PESs will not give
any additional error in the limit of fully expanded fragment PESs. However, this is a limit
we generally have to avoid for efficiency. 
Since higher polynomial
 orders are well known to be important for accuracy, we will use these 
 representations of the PESs for the 
 transformation and accept the inconsistency in the treatment. 
For the expected typical calculation there will, thus, 
be an additional error source from limiting the \chan{mode-coupling expansion}
 when employing auxiliary coordinates. 

\section{Algorithm and implementation \label{sec:impl}}
We report here on a pilot implementation of the double incremental schemes into MidasCpp 
(Molecular interactions, dynamics and simulation Chemistry program package in C++)\cite{midas}.
\chan{
In many aspects, we have made recourse to the most fundamental rather than most
 elaborate schemes. 
For example, our incremental scheme for the evaluation of the electronic energy does 
 not contain any multi-layer approach and we 
 rely on simple hydrogen capping in the fragmentation.
We furthermore 
 generate the PESs on a simple static grid 
 rather than to use adaptive schemes to choose the most important grid points and  
 also the applied transformation algorithm is approximate as previously described.
}
Nevertheless, the implementation provides a proof-of-principle and gives 
 a first insight on the perspective of these schemes for large-scale PES generations. 
Possible beneficial extensions of this scheme for actual 
 applications are discussed in the outlook in Section~\ref{sec:conc}.

\subsection{Partitioning and preparation \label{sec:prep}}
The fragments are chosen on input to the FALCON scheme. 
It is, however possible to obtain this input in a so-called FALCON preparation run, as also 
 used in Ref.~\onlinecite{koni2016}.
The FALCON algorithm is then used to set up the rectilinear, semi-local FALCON coordinates. 
At its end, the FCR is generated, defined by a maximal fragment combination level. 
 We furthermore by default only include those fragment combinations that are 
 covalently bound or direct neighbors and thereby introduce spatially restricted
 FCRs.
Then, we choose for each fragment combination those coordinates that are non-rigid within this 
 mode combination and add hydrogen atoms in case of dangling bonds
 to allow sensible electronic structure calculations on each fragment combination. 
Finally, we generate 
 the auxiliary inter-connecting modes and also these 
 are appropriately capped. 
These steps are described in the following.

\subsubsection{Capping \chan{of} dangling bonds \label{sec:capping}}
The capping of dangling bonds is achieved by replacing the next atom in line 
 with a hydrogen atom and shortening the bond lengths 
 to average bond lengths found in organic compounds from Ref.~\onlinecite{alle1987}.
The chosen bond lengths can be found in Table~S-I in the supplementary material\cite{sm}. 
This treatment does, of course, not allow for cutting of other than single bonds. 
The contribution of each capping atom to the displacement vectors 
 is set to the same value as that of the atom it is bound to. 
By this, we avoid artifacts in the local PESs due to \chan{bond stretches} to the capping 
 atoms. 
These additional contributions to the displacement vectors, of course, 
 lift the orthonormality of the
 basis (if given before). 
\chan{The underlying assumption is that the contribution of the capping atoms to the 
 energy of the displaced structure is similar to that to the reference structure. 
This means we assume that its contribution to the energy difference 
 is small, so that the PES for the respective fragment combination can be 
 approximated by that of the capped system. 
}

\subsubsection{Setting up auxiliary coordinates \label{sec:setup_aux}}

For setting up the auxiliary coordinates for a particular fragment combination, we first 
 identify the space that needs to be spanned by the auxiliary coordinates. 
This space is composed of the 
 inter-connecting modes within the corresponding fragment combination that
 do not correspond to common FALCON modes.
For example,  setting up the auxiliary coordinates for the fragment combination \chan{$\{A,B,C\}$} in the left 
 frame in Figure~\ref{fig:falcon_matrices}, we see that the common FALCON inter-connecting 
 modes of type (A)$\leftrightarrow$(B) only have contributions from atoms in the fragment 
 combination \chan{$\{A,B\}$} and are therefore fully contained in the fragment combination \chan{$\{A,B,C\}$}.
This means that we do not need to set up auxiliary coordinates describing the relative
 motion of (A)$\leftrightarrow$(B).
The remaining inter-connecting modes, that affect \chan{$\{A,B,C\}$},
 are of type  (C)$\leftrightarrow$(D),  (AB)$\leftrightarrow$(CD), and (ABCD)$\leftrightarrow$(EFGH).
Accordingly, they are not fully contained in \chan{$\{A,B,C\}$} and we need to set up the relative 
 motions (AB)$\leftrightarrow$(C) in the auxiliary coordinates.
For this we first initialize the  local translational 
 and rotational coordinates of the relevant fragment combinations.
In the above case, these are (AB) and (C).
These initial coordinates are then modified in a FALCON procedure (described in 
 Ref.~\onlinecite{koni2016}) using the same specifications as in the generation of the
 common FALCON coordinates (only on a fragment combination, that is \chan{$\{A,B,C\}$} in the present example, 
 rather than the complete system).
At the end of the FALCON procedure, reduced Hessians for sets of 
 inter-connecting modes are diagonalized.

The Hessian, which is used during this FALCON procedure to generate auxiliary coordinates for 
 a fragment combination, is that of the full system. 
This
 avoids potential instabilities in the procedure due to 
 our starting structure representing only an equilibrium \chan{structure}
 for the entire system, but not for its 
 fragments and fragment combinations.
The corresponding reduced Hessian can 
 be calculated for the reduced space of the coordinates of interest.
In principle, the same double incremental treatment described here for PESs 
 can also be applied in (semi-numerical) Hessian calculations. 
Making use of such an algorithm to generate the auxiliary coordinates will 
 make this generation of the auxiliary coordinates scale linearly with system size. 
This possibility is, however, not exploited in the present work.
 
The resulting coordinates are then used as auxiliary coordinates for the inter-connecting 
 vibrational modes. 
Finally,  the rotational degrees of freedom of the full fragment combination (\chan{$\{A,B,C\}$} in the present example) 
 are added to the auxiliary coordinates. 
Also these diagonalize the corresponding 
 reduced Hessian of the entire system.

\subsection{Generation of the potential energy surface}
First, the PESs of \chan{all fragment combinations} are calculated separately. 
 This is by default done in the auxiliary basis, if this reduces the number of 
 modes to be considered.
When employing auxiliary coordinates, we have to set their boundaries,
 which define the vibrational space covered in the PES generation, \chan{in a way that fits the
 vibrational space covered by the common FALCON coordinates.} 
In practice, we set the boundaries of auxiliary coordinates ($b^{\rm aux}_{{\rm FC},i}$)
 from the boundaries of the 
 \chan{common} inter-connecting modes ($b^{\rm F}_j$) as, 
\begin{align}
b^{\rm aux}_{{\rm FC},i} = \sum_{j \in \{j_{{\rm max},i}\}} \left| (\mathbf{A}_{\rm FC})_{ij} \right| b^{\rm F}_j,
\end{align}
where $\{j_{{\rm max},i}\}$ contains those $j$s that 
 give
 the $n$ largest contributions $\left| (\mathbf{A}_{\rm FC})_{ij} \right|
  b^{\rm   F}_j$ for the auxiliary coordinate $i$, where $\mathbf{A}_{\rm FC}$ is the 
 transformation matrix.
 $n$ is the smaller of the maximum mode combination level and the number of 
 common FALCON modes 
 to be expressed in the auxiliary coordinates.
\chan{Subsequent to the construction of the fragment combination's PES representation in 
 auxiliary coordinates, it
  is transformed to the representation in common FALCON coordinates, 
 see Section~\ref{sec:app_transform} for details.
After having obtained all fragment combinations' PES representations in common FALCON coordinates,
 the contributions of 
 every fragment combination to the overall PES are calculated according to Eq.~(\ref{eq:incr_bar_combi}) and finally all contributions are summed up using Eq.~(\ref{eq:general_incr_expansion}). 
}  

\subsection{\chan{Coordinate transformation} of polynomial potential energy surfaces \label{sec:app_transform}}

The transformation matrix is set up as described in Section \ref{sec:transform}.
In Appendix \ref{app:trans} we describe our algorithm for transforming 
\chan{a PES in polynomial $n$-mode representation in 
 auxiliary coordinates to a polynomial $n$-mode 
 representation in common coordinates.}
In the usual case where the order of the polynomial expansion is larger than the mode combination 
 level, this transformation procedure can generate higher-order mode combinations
 that have not been included in the original representation of the  PES\cite{yagi2012}.
For sake of computational efficiency, we do not generate these terms.

The applied transformation algorithm scales, in case of a PES representation
  with up to two-mode couplings, 
 cubic with the number of auxiliary coordinates and quadratic with 
 \chan{the number of common FALCON coordinates that are described by auxiliary coordinates 
 in the original PES representation.
The algorithm employed here} is restricted to PES representations of polynomial form. 
Other algorithms are possible, which allow the usage of other analytic representations of the 
 PES.
It is for example possible to estimate the energy corresponding to displacements along the 
 common FALCON coordinates from
 the PES \chan{representation} in auxiliary coordinates by interpolation and then fit a new   
 representation of the PES to \chan{these} points. 
Since, the exact points are typically not explicitly contained in 
 the auxiliary PES representation, 
 interpolation techniques using derivative information 
 (as, e.g., used in  Refs.~\onlinecite{mati2009,spar2010})
 will improve the description further in this case. 
These possibilities are, however, not a subject of the work presented here.

\section{Setup and computational details \label{sec:cd}}

The generation of the common and auxiliary FALCON coordinates \chan{and the PES 
 construction require} the accessibility of electronic energies \chan{and gradients}.
Most electronic structure calculations are performed with the {\sc Orca} program system\cite{orca} 
 employing the {\emph{Hartree--Fock with three corrections} (HF-3c)
  method\cite{sure2013}.
We performed additional calculations using density functional methods, 
 more precisely the Becke--Perdew functional (BP86) \cite{beck1988,perd1986} and a split valence polarization 
 (SVP) double-$\zeta$ basis set \cite{scha1992}
 using the {\sc Turbomole}\cite{ahlr1989,furc2014,TURBOMOLE510} program package.
In these calculations, we applied the resolution-of-the-identity approximation (RI) with the
 corresponding auxiliary basis set \cite{eich1995}.

The initial structures of the oligo-phenyl test systems are obtained with  
 Avogadro\cite{avogadro}. 
The structures \chan{have then been} optimized using the respective electronic structure method.
After defining the initial groups for the FALCON procedure
 to be the single phenyl rings, we determine the common FALCON coordinates
 applying the following settings, see Ref. \onlinecite{koni2016} for details.
We apply distance-based coupling estimates and 
 a degeneracy threshold of 1$\cdot 10^{-5}$~bohr$^{-1}$.
Relaxation is included for each initial group, i.e., each fragment, meaning that the reduced 
 Hessian for the coordinates for each group/fragment is diagonalized separately.
We furthermore relax the inter-connecting coordinates separately for each type of coordinates. 
In this relaxation setup, only coordinates with contributions from the same fragments 
  are mixed and therefore 
 the overall local character (as exemplified in Figure~\ref{fig:falcon_matrices}) is maintained.  

The fragments for the incremental PES generations are the same as the initial 
 groups used in the FALCON procedure. 
The applied FCRs only contain direct neighbors, with any atom pair closer
 than 3.5~bohr. 
We furthermore employ effective FCRs, omitting all fragment combinations 
 with overall zero contribution.

We have applied the 
 existing functionality in MidasCpp\cite{midas} 
 for grid-based $n$-mode expan\-sions\cite{kong2006,toff2007}
 for the generation of the PESs of the fragments and fragment combinations.
Therefore, we
have applied equidistant grids with boundaries chosen to be 
 the classical turning points corresponding
 to a harmonic oscillator with quantum number of $v=10$.
These turning points are calculated using the quasi-harmonic FALCON frequencies of the
 respective modes.
For the one-mode grids, we have applied 16 grid points per dimension, whereas we have used 
  eight grid points per dimension, meaning 64 grid points for each two-mode cut. 
All PESs are approximated by an $n$-mode expansion up to two-mode couplings
 with the cut functions fitted to polynomials.
The \chan{maximum} polynomial order of the PES fits is in all cases set to twelve.

In the following, we apply the  short-hand notations for the differently 
 obtained PES representations:
We denote full $n$-mode expansion by ``$n$M'' (only 2M in the present work) and use 
  ``DIF-$f$F$n$M'' and 
 ``DIFACT-$f$F$n$M'' for DIF and DIFACT expansion, respectively, up to $f$th order in fragment 
 combination and $n$th order in mode combination.
As mentioned above, the FCR is by default spatially restricted, i.e., 
 we consider only neighbor couplings.   

For the validation of the PES we have performed state-specific vibrational 
 self-consistent field (VSCF) calculations for the one-mode excited states. 
These calculations were performed with the VSCF implementation\cite{hans2010} in 
 MidasCpp\cite{midas}, using a b-spline basis\cite{toff2010} of the order ten with 
 a b-spline density of 0.8 in the region explicitly covered in the PES
  generation, i.e., up to the classical turning point with the harmonic quantum number 
  $v=10$.
Again these classical turning points were 
  calculated from the quasi-harmonic frequency of the respective FALCON modes. 

\section{Numerical examples for oligo-phenyls \label{sec:tech_test} }

For first numerical examples for the double incremental scheme, 
 we have chosen chain-like tetra- and hexa-phenyl molecules.
We employ structures with alternating dihedral angles between the phenyl entities of about $\pm$30 degrees
 (see Figure~\ref{fig:struc_oligo-phenyl}).
These structures represent stationary points for the chosen method to calculate the electronic 
 energy.
\begin{figure} 
\includegraphics[width=\textwidth]{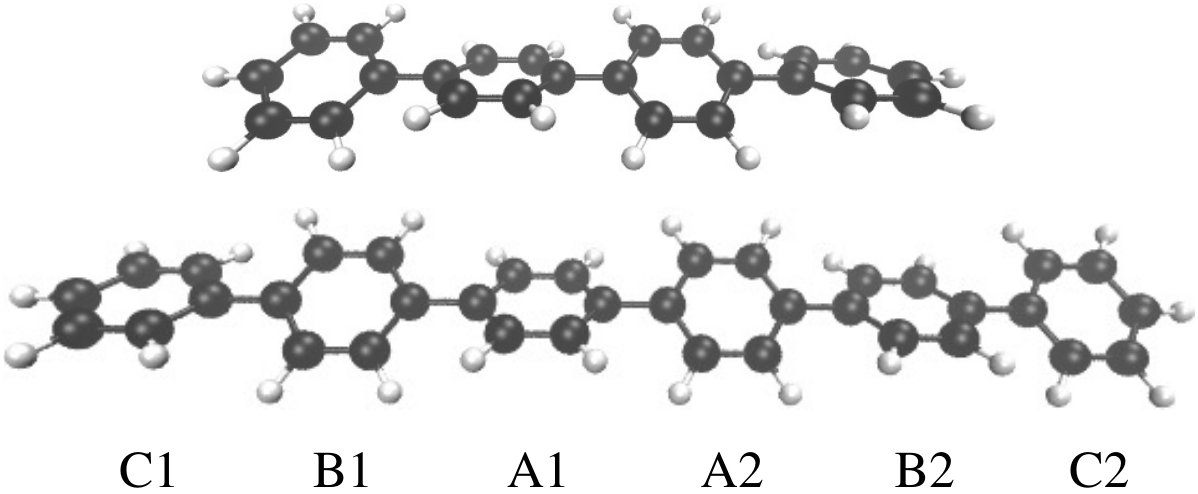} 
\caption{
Applied structure for tetra- and hexa-phenyl (HF-3c).
\label{fig:struc_oligo-phenyl}
}
\end{figure} 
We treat every phenyl entity as an individual fragment and initial group of the 
 FALCON algorithm.
We apply inter-connecting FALCON coordinates for the tetra-phenyl of kind [B1]-(A1)$\leftrightarrow$(A2)-[B2] 
 and (B1)$\leftrightarrow$(A1-A2)$\leftrightarrow$(B2) and that for the hexa-phenyl are obtained as 
 [C1-B1]-(A1)$\leftrightarrow$(A2)-[B2-C2], 
 [C1]-(B1)$\leftrightarrow$(A1-A2)$\leftrightarrow$(B2)-[C2], and 
 (C1)$\leftrightarrow$(B1-A1-A2-B2)$\leftrightarrow$(C2).  
The naming of the fragments is according to Figure~\ref{fig:struc_oligo-phenyl},
and the nomenclature for the classification of the different 
 types of modes that of section \ref{sec:semi-local-coord}.

\chan{
The errors in PES representations due to different approximations 
  are assessed by the difference in the  
state-specific VSCF energy for the fundamentally excited states from the VSCF zero-point energy.
}

\subsection{Effective fragment combination ranges and computational cost}

In the FCR, we only account for direct neighbors.
The effective FCRs, i.e., those without zero contribution according
 to Eq.~(\ref{eq:coef}) are 
 summarized in Table~\ref{tab:effectiveFCR}.
\begin{table}
\caption{
Spatially restricted effective FCRs considering only neighbor couplings 
  for tetra- and hexa-phenyl, with 
 a maximal  fragment combination level $(f)$ from two to four.
\label{tab:effectiveFCR}
}
\begin{tabular}{ll}
\hline
\hline
 $f$ & Effective FCR \\
\hline
\multicolumn{2}{l}{Tetra-phenyl}\\
\hline
 2   & $ \{ \{A1\}, \{A2\}, \{B1,A1\}, \{A1,A2\}, \{A2,B2\}\}$\\
 3   & $ \{ \{A1,A2\}, \{B1,A1,A2\}, \{A1,A2,B2\}\}$\\
 4   & $ \{\{B1,A1,A2,B2\}\} $ \\
\hline
\multicolumn{2}{l}{Hexa-phenyl}\\
\hline
 2   & $ \{ \{B1\}, \{A1\}, \{A2\}, \{B2\}, \{C1,B1\}, \{B1,A1\}, \{A1,A2\}, \{A2,B2\}, \{B2,C2\} \}$  \\
 3   & $ \{  \{B1,A1\}, \{A1,A2\}, \{A2,B2\},  
      \{C1,B1,A1\}, \{B1,A1,A2\}, \{A1,A2,B2\}, \{A2,B2,C2\}\}$  \\
 4   & $ \{  \{B1,A1,A2\}, \{A1,A2,B2\},  
       \{C1,B1,A1,A2\}, \{B1,A1,A2,B2\}, \{A1,A2,B2,C2\}\}$   \\
\hline
\hline
\end{tabular}
\end{table}
\chan{
In these chain-like examples with spatially restricted 
 FCR,  a significant number of lower-order fragment combinations have zero
 contribution to the overall expansion, as shown more generally in Appendix \ref{app:chain-eff}}.
This also affects the required number of SPCs and accumulated cost of all SPCs.
\begin{figure}
\includegraphics[width=0.8\textwidth]{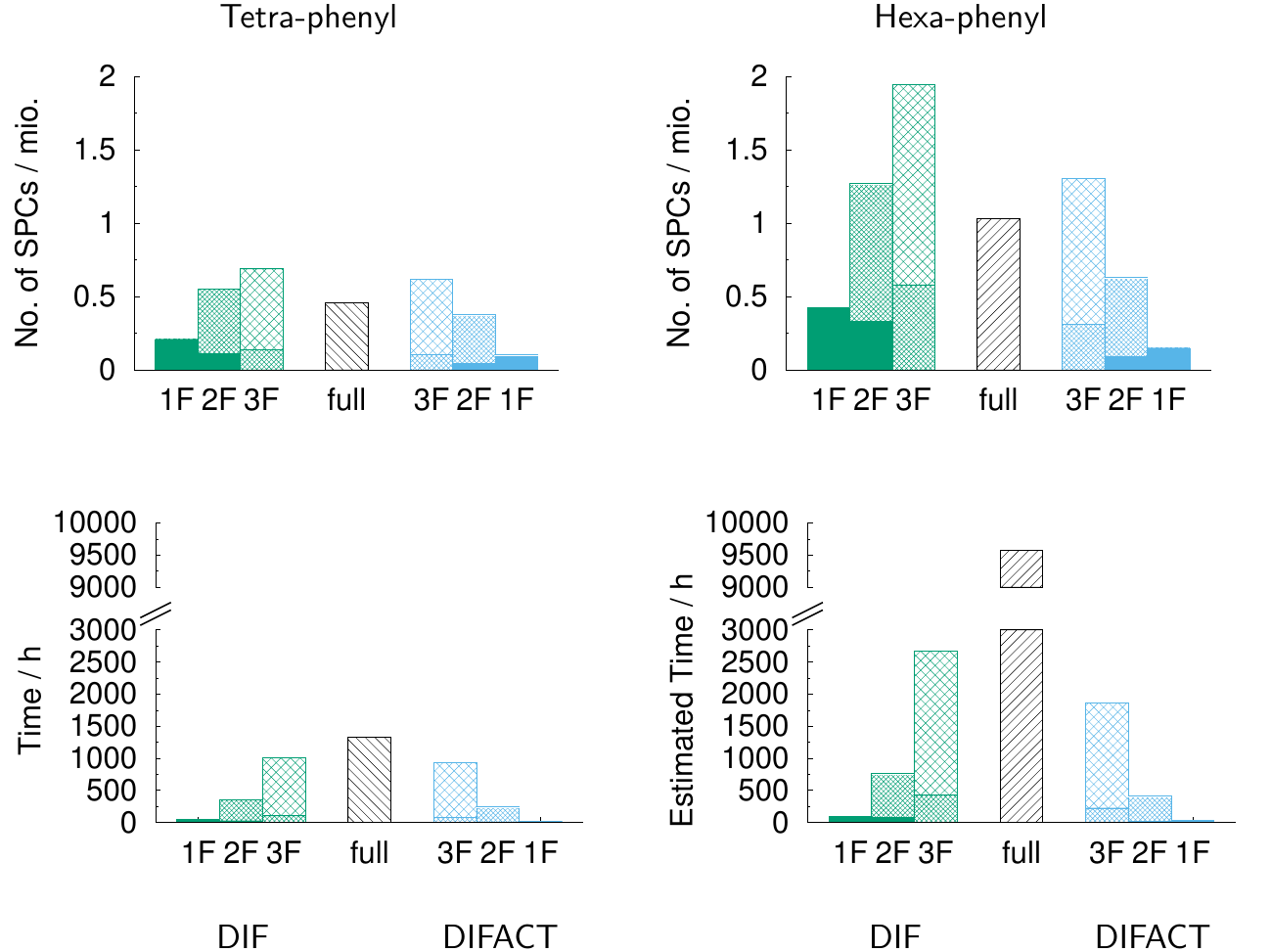}
\caption{
Number of required SPCs and accumulated wall time of 
 SPCs (HF-3c, on Intel Ivy Bridge 20 cores @ 2.8 GHz / 128GB; Dell C6220)
 for different spatially restricted effective FCRs for tetra-phenyl and hexa-phenyl
 in full vibrational space. 
The different patterns indicate, whether the contributions stem from 
 a one-fragment potential (filled), a two-fragment potential (fine criss-cross), a three-fragment potential (coarse criss-cross), or the potential for the 
 full four- or six-fragment system (feint ruled). 
The wall times for hexa-phenyl are estimated (see main text).
\label{fig:sp_timing}
}
\end{figure}
These computational figures are shown in Figure~\ref{fig:sp_timing} for different 
 schemes to generate the PES.
As expected from the discussion in Section~\ref{sec:scaling}, the number of required 
 SPCs increases significantly with the fragment combination level.
It is  generally larger
 for the DIF model than for the DIFACT approach with the same fragment combination level.
The difference between these two models 
 is, as expected, significantly larger for the hexa-phenyl than for the tetra-phenyl example.
\chan{
The number of required SPCs in 
 double incremental PES constructions 
 is in many cases
 larger than the required number of SPCs for generating the  
 full 2M PES.}

To give an impression on how the computational cost differs between the different approaches, we have
 timed the individual SPCs on nodes of the type 
 Intel Ivy Bridge 20 cores @ 2.8 GHz / 128GB; Dell C6220. 
The times plotted for the hexa-phenyl are estimated from the number of SPCs and 
 averaged wall time for several SPCs of displaced structures of the respective molecules. 
This means, that the timings reported in the bottom part of Figure~\ref{fig:sp_timing} 
 should be considered ballpark figures rather than rigorous 
 assessments. 
They can still give valuable insight into the computational gains in the double incremental scheme in 
 actual computations. 
First of all, we see  that
 all shown double incremental expansions require less computational resources than the 
 2M reference calculations. 
As expected, this difference is more pronounced for the hexa-phenyl than for the tetra-phenyl example. 
The accumulated computational cost for SPCs 
  is typically the bottleneck in PES generations.
The double incremental approaches, therefore, 
 offer a large gain in the feasibility of PES generations for large molecular 
 systems.
The relative gain is  expected to be more beneficial the larger the investigated system is (see also Section~\ref{sec:scaling}).

\subsection{Error assessment}

In this section, we assess how well we can reproduce the 
 conventional, non-incremented calculation 
 with the different approximate double incremental schemes in FALCON coordinates.
The errors to be assessed are (i) the error introduced by the truncated double incremental expansion, 
 including the capping of dangling bonds 
 and (ii) a transformation error, in case \chan{auxiliary modes are employed}.

\subsubsection{Error assessment for inter-connecting modes}

For the first error assessment, we restrict the discussion to 
 the vibrational subspace of inter-connecting 
 modes in the tetra- and hexa-phenyl examples. 
For these modes, the errors introduced by the double incremental expansion 
 and usage of auxiliary coordinates are expected to be particularly large, 
 since (i) these modes include often more fragments than covered in the largest fragment combination 
 level and (ii) the transformation procedure is only applied to these modes. 

Table \ref{tab:tetra-phenyl} shows the results obtained considering only these inter-connecting 
 modes for tetra-phenyl \chan{and employing} HF-3c \chan{for the SPCs}.
\begin{table}
\caption{
Zero-point energies (ZPEs)
 and differences of the fundamental energies 
 from the ZPE ($\Delta E$) obtained from state-specific VSCF \chan{calculations} for tetra-phenyl
 for different PES representations (HF-3c)  and considering only inter-connecting modes. 
$\omega$ are the corresponding quasi-harmonic FALCON frequencies. 
All energies are given in {\rcm}.
The type label IC1 refers to [B1]-(A1)$\leftrightarrow$(A2)-[B2] 
 and IC2 to (B1)$\leftrightarrow$(A1-A2)$\leftrightarrow$(B2) 
 type of FALCON coordinate.
\label{tab:tetra-phenyl} 
}
\begin{tabular}{lc
@{\extracolsep{3pt}}
rrrrrrrrrr
@{\extracolsep{0pt}}
}
\hline           
\hline
          & Type & $\omega$           &  {2M}                  &  \multicolumn{2}{c}{DIF-$f$F2M}                  & \multicolumn{2}{c}{DIFACT-$f$F2M}     \\
\cline{5-6} 
\cline{7-8} 
$f$        &      &                    &                        & 2                     & 3                       & 2                       & 3                      \\
\hline
ZPE       &      &                    & 2818.97 & 2817.01 & 2818.92 & 2817.70 & 2819.15\\
\hline                                                                                        
\multicolumn{6}{l}{Fundamental excitation energies} \\                                                                 
\hline                                                                                        
1         & IC2    & 45.95 & 56.96 & 57.03 & 56.96 & 58.01 &  56.89 \\
2         & IC2    & 59.23 & 87.30 & 87.49 & 87.32 & 85.68 &  89.40 \\
3         & IC2    & 74.54 & 82.12 & 82.27 & 82.12 & 82.57 &  82.49 \\
4         & IC2    & 83.84 & 102.26& 102.63& 102.26& 102.63&  100.15\\
5         & IC2    & 87.92 & 97.71 & 97.67 & 97.70 & 98.73 &  97.84 \\
6         & IC1    & 102.51& 114.85& 115.43& 114.87& 116.89&  115.72\\
7         & IC2    & 132.33& 138.46& 137.96& 138.45& 138.16&  138.65\\
8         & IC1    & 326.79& 329.03& 329.30& 329.03& 329.81&  329.29\\
9         & IC2    & 349.13& 352.29& 352.53& 352.28& 352.54&  352.03\\
10        & IC1    & 370.57& 371.80& 371.15& 371.77& 370.95&  371.97\\
11        & IC2    & 376.65& 374.75& 374.55& 374.75& 374.53&  374.76\\
12        & IC2    & 401.39& 403.34& 404.43& 403.39& 404.70&  403.15\\
13        & IC2    & 428.26& 429.10& 428.76& 429.07& 429.02&  429.02\\
14        & IC2    & 499.72& 499.83& 499.48& 499.83& 499.75&  499.67\\
15        & IC1    & 503.33& 504.05& 503.27& 504.07& 503.29&  504.23\\
16        & IC2    & 534.47& 531.13& 530.82& 531.10& 530.72&  531.09\\
17        & IC1    & 549.26& 549.31& 549.10& 549.30& 548.91&  549.37\\
18        & IC1    & 658.01& 654.06& 652.75& 654.01& 652.71&  654.02\\
\hline                                                                                                            
\multicolumn{6}{l}{$\Delta$ ZPE to} \\                                                                 
\hline                                                                                                            
2M       &       &                    &                        & $-$1.96 & $-$0.05 & $-$1.27& 0.19\\
DIF-$f$F2M &       &                    &                        &                        &                          &  0.69& 0.24\\
\hline                                                                                                            
\multicolumn{6}{l}{Root mean square deviation of $\Delta E$ to  } \\                                                                 
\hline                                                                                                            
2M       &       &                   &                         & 0.54&   0.02&  0.93&  0.75\\
DIF-$f$F2M    &       &                   &                         &                        &                          &  0.67&  0.74\\
\hline                                                                                                                                                             
\multicolumn{6}{l}{Maximum absolute deviation of $\Delta E$ from}  \\                                                                                            
\hline                              
2M     &       &                   &                         & 1.31&   0.05&  2.04&  2.11\\
DIF-$f$F2M     &       &                   &                         &                        &                          & 1.81&  2.11\\
\hline
\hline
\end{tabular}
\end{table}
First we assess the capping and fragmentation error (i): 
We see that the root mean square deviation (RMSD) 
 of the fundamental VSCF excitation energies obtained 
 with a DIF-2F2M potential  
 compared to the 2M reference amounts to only 
 0.54~{\rcm} and the maximum absolute deviation is 1.31~{\rcm}.
The results for the DIF-3F2M potential and the reference potential are in almost
 perfect agreement with a maximum absolute deviation \chan{in 
 fundamental excitation energies} of 0.05~{\rcm}. 
The absolute \chan{deviations} of the VSCF zero-point energy obtained with 
 \chan{DIF-2F2M and DIF-3F2M, respectively, are} only slightly 
 larger than  maximum absolute deviation of \chan{the excitation energies}. 
It
 amounts to 1.96~{\rcm} for DIF-2F2M and to 0.05~{\rcm} for DIF-3F2M.
These findings suggest very good and quick convergence of the fragmentation error with increasing 
 fragment combination level. 

Next, we turn to the transformation error (ii), i.e., the deviation between the fundamental 
 excitation energies obtained with a {PES generated} in auxiliary coordinates 
 from those for the corresponding PES representation without this approximation.
We see that the RMSD caused by this error lies around 0.7~{\rcm} both {for} 
 DIFACT-2F2M and DIFACT-3F2M and the maximum absolute deviation 
 is at 1.81~{\rcm}} for DIFACT-2F2M and 
 at 2.11~{\rcm} for DIFACT-3F2M. 

Very similar, though slightly larger errors \chan{caused by} the fragmentation and transformation 
 are obtained for the larger hexa-phenyl molecule (see Table~ \ref{tab:hexa-phenyl}
 and Table~S-II in the supplementary material\cite{sm} for full details), where we 
 again consider only inter-connecting modes. 
\begin{table}
\caption{
Zero-point energies (ZPEs) as well as root mean square deviation  
 and maximum absolute deviation of the difference in
 VSCF state energies for singly excited states from the ZPE ($\Delta E$) for   
 hexa-phenyl for different PES representations (HF-3c)
 and considering only inter-connecting modes. 
 All energies are given in {\rcm}.
\label{tab:hexa-phenyl}
}
\begin{tabular}{lc
@{\extracolsep{3pt}}
rrrrrrrr
@{\extracolsep{0pt}}
}
\hline           
\hline
          & &  &{2M}    &  \multicolumn{3}{c}{DIF-$f$F2M}                                                & \multicolumn{3}{c}{DIFACT-$f$F2M}     \\
\cline{5-7} 
\cline{8-10} 
$f$          &      &                    &                        & 2                     & 3                       & 4                      & 2                       & 3                      & 4      \\
\hline                                               
ZPE       &      &                    & 4797.68 & 4794.19 &  4797.55 & 4797.67 & 4796.02 & 4798.16 & 4797.62\\
\hline                                                                                                                                                                                           
\multicolumn{10}{l}{$\Delta$ ZPE to } \\                                                                 
\hline
\multicolumn{3}{l}{2M}     &                          & $-$3.49& $-$0.12& $-$0.00 & $-$1.66&  0.48 & $-$0.06\\
\multicolumn{3}{l}{DIF-$f$F2M}   &                        &                        &                          &                         &  1.83&  0.60& $-$0.06\\
\hline                                                          
\multicolumn{10}{l}{Root mean square deviation of $\Delta E$ to } \\   
\hline                                                          
\multicolumn{3}{l}{2M}     &                         & 0.80&  0.04&   0.00&  1.25&  0.88& 0.66\\
\multicolumn{3}{l}{DIF-$f$F2M}    &                         &                        &                          &                         &  0.75&  0.88& 0.66\\
\hline                                                                     
\multicolumn{10}{l}{Maximum absolute deviation of $\Delta E$ from }  \\           
\hline                              
\multicolumn{3}{l}{2M}    &                         &  2.11&  0.13&  0.01&  3.55& 2.96& 2.82\\
\multicolumn{3}{l}{DIF-$f$F2M}    &                         &                       &                           &                         &   2.13& 2.92& 2.81\\
\hline
\hline
\end{tabular}
\end{table}
\chan{
All in all, we see also here quick convergence of the double incremental scheme} with 
 almost perfect agreement already for the DIF-3F2M potential with a maximum absolute deviation  
 in the fundamental excitation energies of 0.13~{\rcm}. 
Concerning the transformation error, we obtain RMSDs between 0.66 and 0.88~{\rcm} 
 for fragment combinations levels of two to four. 
These errors are, hence,
 rather close to those for the tetra-phenyl example with only inter-connecting modes
 (\chan{see Table~\ref{tab:tetra-phenyl}}).

\subsubsection{Error assessment in full space}

Considering all 120 vibrational degrees of freedom of the tetra-phenyl example, we obtain 
 very similar fragmentation errors compared to the above case with only inter-connecting modes 
(compare Tables \ref{tab:tetra-phenyl} and \ref{tab:4-phenyl_all_hf-3c}, see 
 also S-III in the supplementary material\cite{sm} for all 120 fundamental excitation energies 
 for tetra-phenyl using PESs obtained with HF-3c).
\begin{table}
\caption{Zero-point energies (ZPEs) as well as root mean square deviation (RMSD) 
 and maximum absolute deviation (MAX) of the difference in
 VSCF state energies for singly excited states from the ZPE ($\Delta E$) for  
 tetra-phenyl for different PES representations (HF-3c). 
The deviations for inter-connecting (IC) modes and intra-fragment (INTRA) modes are shown separately and combined (All).
All energies are given in {\rcm}.
\label{tab:4-phenyl_all_hf-3c}} 
\begin{tabular}
{ll
@{\extracolsep{2pt}}
rrrr
r
rrr
@{\extracolsep{0pt}}
}
\hline
\hline
 &  &  &   {2M} & \multicolumn{2}{c}{DIF-$f$F2M}   & \multicolumn{2}{c}{DIFACT-$f$F2M}\\
\cline{5-6}
\cline{7-8}
$f$   &      &          &       & 2          & 3         & 2          & 3\\
\hline
\multicolumn{2}{l}{ZPE}
 &              &  86266.64 &  86263.37 & 86266.54 &  86263.54&  86267.29 \\
\hline
\multicolumn{8}{l}{$\Delta$ ZPE to 2M}\\
\hline
      &      &         &                              &  $-$3.27& $-$0.10 & $-$3.10&  0.65\\
\hline
\multicolumn{8}{l}{$\Delta$ ZPE to DIF-$f$F2M}\\
\hline
      &      &         &                              &                         &                         & 0.17&  0.76\\
\hline                                                          
\multicolumn{8}{l}{RMSD of $\Delta E$ to 2M} \\   
\hline                                                          
\multicolumn{3}{l}{All}     &  & 0.56 & 0.02 & 0.76 & 0.59\\
\multicolumn{3}{l}{IC}      &  & 0.51 & 0.02 & 1.10 & 1.51\\ 
\multicolumn{3}{l}{INTRA}   &  & 0.57 & 0.02 & 0.69 & 0.12\\
\hline
\multicolumn{8}{l}{RMSD of $\Delta E$ to DIF-$f$F2M} \\   
\hline                                                          
\multicolumn{3}{l}{All}      & &             &             & 0.44 & 0.59\\
\multicolumn{3}{l}{IC}       & &             &             & 0.82 & 1.51\\
\multicolumn{3}{l}{INTRA}    & &             &             & 0.32 & 0.12\\
\hline
\multicolumn{8}{l}{MAX of $\Delta E$ from 2M  } \\   
\hline                                                          
\multicolumn{3}{l}{IC}       & & 1.37 & 0.05 & 2.12 & 4.79 \\
\multicolumn{3}{l}{INTRA}    & & 2.35 & 0.09 & 2.65 & 0.29\\
\hline                      
\multicolumn{8}{l}{MAX  of $\Delta E$ from DIF-$f$F2M } \\   
\hline                                                          
\multicolumn{3}{l}{IC}       & &             &             & 1.79 & 4.79 \\
\multicolumn{3}{l}{INTRA}    & &             &             & 0.57 & 0.28\\
\hline
\hline
\end{tabular}
\end{table}
The deviations found for the zero-point energy are again larger than those for the
 fundamental excitation energies. 
Both decrease systematically when increasing the maximal order of fragment combination 
 in the DIF scheme. 
The maximum absolute deviation of the energy differences obtained with DIF-2F2M 
 from the 2M reference corresponds to an intra-fragment mode
 and amounts to 2.35~{\rcm}.
The maximum absolute deviation for the inter-connecting modes is 1.37~{\rcm} in \chan{this} case.
For DIF-3F2M, all fundamental excitation energies agree within $\pm$~0.1~{\rcm}.
As expected, the transformation error is generally larger for the directly 
 affected inter-connecting modes than for the intra-fragment modes. 
The latter are only indirectly affected by the transformation via the coupling elements 
 to the inter-connecting modes.
In the present case, the transformation error decreases for intra-fragment modes  
 and increases for inter-connecting modes when going from DIFACT-2F2M to DIFACT-3F2M. 

\begin{figure}
\includegraphics[width=\textwidth]{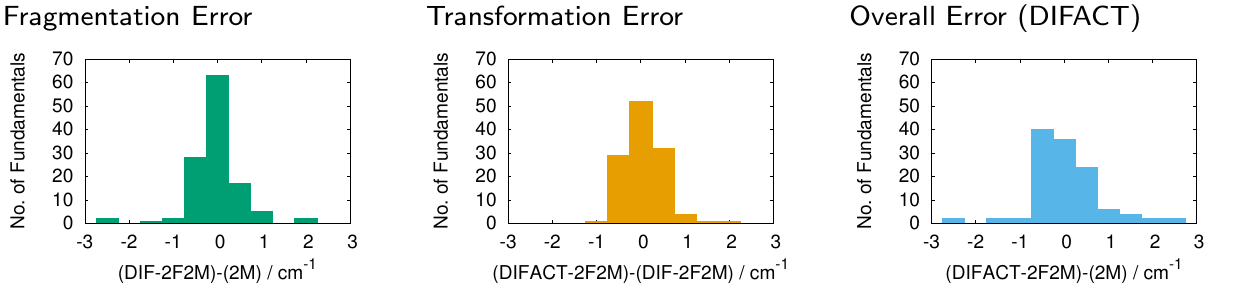}
\caption{
Error distributions for the VSCF fundamental \chan{excitation} energies for DIF-2F2M and DIFACT-2F2M
 representations of the PES (HF-3c)  
 for tetra-phenyl in full vibrational space with respect to each other (middle graph) 
 and the reference results for the 2M PES representation (left and right graph, respectively).
\label{fig:error_distributions}
}
\end{figure}

Figure~\ref{fig:error_distributions} shows the error distributions for 
 fragmentation, transformation, and overall error for the example of 
 PES \chan{representations} up to second order in fragment combination. 
The error distributions are well centered on zero.
The overall error has similar contributions from the fragmentation and transformation 
 error in this case, as has already been suggested by the similar 
 RMSDs.
This supports that the transformation at the 2F level does not lead to an inherent larger 
 error for the full procedure.
An analogous analysis for the 3F case (not shown) yields an almost perfect
 agreement of 
 the DIF-3F2M results with the 2M reference results and also the transformation 
 error for most fundamental \chan{excitation} energies, lies close to zero:   
Only three fundamental VSCF \chan{excitation} energies 
 obtained with a DIFACT-3F2M representation of the PES
 deviate by more than 0.6~{\rcm} from 
 the DIF-3F2M results. 
Those have an absolute deviation of 3.71~{\rcm}, 1.88~{\rcm}, and 4.79~{\rcm} and 
 correspond to vibrations of the type 
 (B1)$\leftrightarrow$(A1-A2)$\leftrightarrow$(B2)  with
 quasi-harmonic frequencies clearly below 100~{\rcm}. 
This means that the comparatively 
 large maximum absolute transformation error of 4.79~{\rcm} for DIFACT-3F2M is
 likely an outlier. 
This suggests, that only these three common FALCON coordinates are not very well
 resembled in the auxiliary coordinates of the fragment combinations in the present case,
 while the PES contributions of all others are surprisingly well maintained 
 in the potential transformations.

Next, we turn to the hexa-phenyl example\chan{, whose results are summarized in Table~\ref{tab:6-phenyl_all_hf-3c}}.
\begin{table}
\caption{Zero-point energies (ZPEs) as well as root mean square deviation (RMSD) 
 and maximum absolute deviation (MAX) of the difference in
 VSCF state energies for singly excited states from the ZPE ($\Delta E$) for  
 hexa-phenyl for different PES representations (HF-3c). 
The deviations for inter-connecting (IC) modes and intra-fragment (INTRA) modes are shown separately and combined (All).
 \label{tab:6-phenyl_all_hf-3c}} 
\begin{tabular}{ll
@{\extracolsep{4pt}}
rrrr
rrrr
@{\extracolsep{0pt}}
}
\hline
\hline
            &               &                   &        {2M}           & \multicolumn{2}{c}{DIF-$f$F2M}& \multicolumn{2}{c}{DIFACT-$f$F2M}\\
\cline{5-6}
\cline{7-8}
$f$   &      &          &       & 2          & 3         & 2          & 3\\
\hline
\multicolumn{8}{l}{ZPE}\\
\hline
   &      &          & 127147.62 & 127140.68 & 127147.33 & 127142.81 & 127147.80\\
\hline
\multicolumn{8}{l}{$\Delta$ ZPE to 2M}\\
\hline
   &      &          &       & $-$6.94 & $-$0.29 & $-$4.81 & 0.17\\ 
\hline
\multicolumn{8}{l}{$\Delta$ ZPE to DIF-$f$F2M}\\
\hline
   &      &          &       &                           &                                & 2.13 & 0.47\\
\hline
\multicolumn{8}{l}{RMSD of $\Delta E$ to 2M} \\   
\hline                                                          
\multicolumn{3}{l}{All}   &   & 0.75  & 0.04   &  0.97    & 0.65  &  \\
\multicolumn{3}{l}{IC}    &   & 0.77  & 0.04   &  1.46    & 1.54  &  \\ 
\multicolumn{3}{l}{INTRA} &   & 0.75  & 0.04   &  0.84    & 0.17  &  \\
\hline
\multicolumn{8}{l}{RMSD of $\Delta E$ from DIF-$f$F2M} \\   
\hline                                                          
\multicolumn{3}{l}{All}   &   &             &             & 0.49   & 0.65     \\
\multicolumn{3}{l}{IC}    &   &             &             & 0.99   & 1.54     \\
\multicolumn{3}{l}{INTRA} &   &             &             & 0.30   & 0.16      \\
\hline
\multicolumn{8}{l}{MAX of $\Delta E$ from 2M } \\   
\hline                                                          
\multicolumn{3}{l}{IC}    &   & 2.17  & 0.13  & 4.34  & 4.83  \\
\multicolumn{3}{l}{INTRA} &   & 3.13  & 0.22  & 3.12  & 0.45 \\
\hline                                       
\multicolumn{8}{l}{MAX of $\Delta E$ from DIF-$f$F2M } \\   
\hline                                                          
\multicolumn{3}{l}{IC}    &   &             &             & 3.14  & 4.83  \\
\multicolumn{3}{l}{INTRA} &   &             &             & 0.65  & 0.42  \\
\hline
\hline
\end{tabular}
\end{table}
\chan{The full data set of calculated excitation energies for hexa-phenyl can be 
 found in Table~S-IV in the supplementary material\cite{sm}.
Again the largest deviation of $-$6.94~{\rcm} 
 is obtained for the zero-point energy when including fragment combinations 
 up to second order. 
The deviations in the fundamental excitation energies obtained in the 
  double incremental schemes from the 2M reference results  
 for hexa-phenyl are of similar size as those for the tetra-phenyl discussed above 
 (see Table~\ref{tab:4-phenyl_all_hf-3c}), though slightly larger. 
The DIF-3F2M results agree again very well with the reference results, with 
 a maximum deviation in fundamental excitation energies of 0.22~{\rcm} (see Table V).
Again, 
the majority of the fundamental excitation energies have transformation errors 
 below 1~{\rcm} and 
 only few fundamental excitation energies, which all correspond to low-lying inter-connecting modes,
 exhibit a considerably larger  transformation error.
The maximum absolute deviation due to the potential transformation amounts to 4.83~{\rcm} in this case.
}

We have, thus, found in this first numerical study 
that the double incremental scheme can indeed yield very 
 good agreement with the supermolecular calculations. 
Moreover, the convergence with respect to fragment combination order is rather quick and already the
 results for the three-fragment combinations are almost perfectly in line with the reference values.
 The results illustrate that the semi-local FALCON coordinates provide a good set of coordinates that
 allows distance cutoff in the \chan{PES} construction with fast convergence. 
 The controlled and well-defined locality of the FALCON coordinates is essential for the efficiency of DIF procedure.
 
We see that the transformation error is somewhat similar in all cases. 
It is generally lower for intra-fragment modes than for inter-connecting modes. 
There is no clear tendency for the transformation error, when going to 
 higher-order fragment combinations. However, this is also what should be expected.
 What is required to reduce the transformation errors is a better transformation such as
 a better algorithm and/or a higher level in mode coupling. Thus, a 3M expansion and transformation would 
 allow the transformed potential to capture better the correct features of the true potential. 
However, already for the 2M case, we observe a transformation error of less than 5~{\rcm} in all fundamental energies, 
and the majority of cases have errors less than 1~{\rcm}. These errors are modest,  both
 in view of the approximations involved in the present transformation algorithm applied
 in the DIFACT scheme as described earlier, but also in the sense of being modest compared to other    
 sources of errors,
 including the 2M restriction, errors from the electronic structure methodology applied, etc.
The good performance \chan{in the PES transformation} can, at least partly, be rationalized by the
very similar fashion in which the common and auxiliary modes are set up. 
In this way, we generate auxiliary coordinates that in many cases resemble the 
 common modes to a large 
 extent, and thereby only modest coordinate rotations take place in the potential transformation. 
Modest coordinate rotations are expected to lead to  higher 
accuracy in the current potential transformation algorithm. 
Other algorithms may be less sensitive with respect to this aspect. 
Recalling that this additional transformation enables linear scaling of \chan{the
 accumulated computational cost of the SPCs in} the PES generation,
 this error seems legitimate, especially when interested in excitations dominated 
 by intra-fragment modes. 
Still if one needs more accurate potentials, 
 the DIF scheme also offers a significant reduction of the 
 computational cost compared to the complete calculation and has the advantage 
 of clear and rigorous limits towards 
 the full expansion.

\subsubsection{Change of electronic structure method}
In Sections V B 1 -- V B 2, we have seen that the error that we introduce
  by fragmentation in the double incremental scheme lies, even for the 
 DIFACT-2F2M approximation, below 5~{\rcm} in all cases and is thereby  
 below the error made through other approximations, such 
 as the electronic structure method.
This suggests, that it is likely beneficial to use a more reliable (and thereby more expensive) electronic structure 
 method. 
\chan{Here, we illustrate that the DIFACT scheme indeed makes it feasible 
 to use more accurate electronic structure methods in the  
 PES construction with more than hundred degrees of freedom. 
We compare the DIFACT-2F2M potentials for tetra-phenyl obtained with the HF-3c to that from RI-BP86/SVP SPCs.}
The respective stick spectra are together with a reference 2M spectrum for HF-3c shown in Figure~\ref{fig:bp86}.
The respective data can be found in Tables S-III and S-V in the supplementary material\cite{sm}.
\begin{figure}
\includegraphics[]{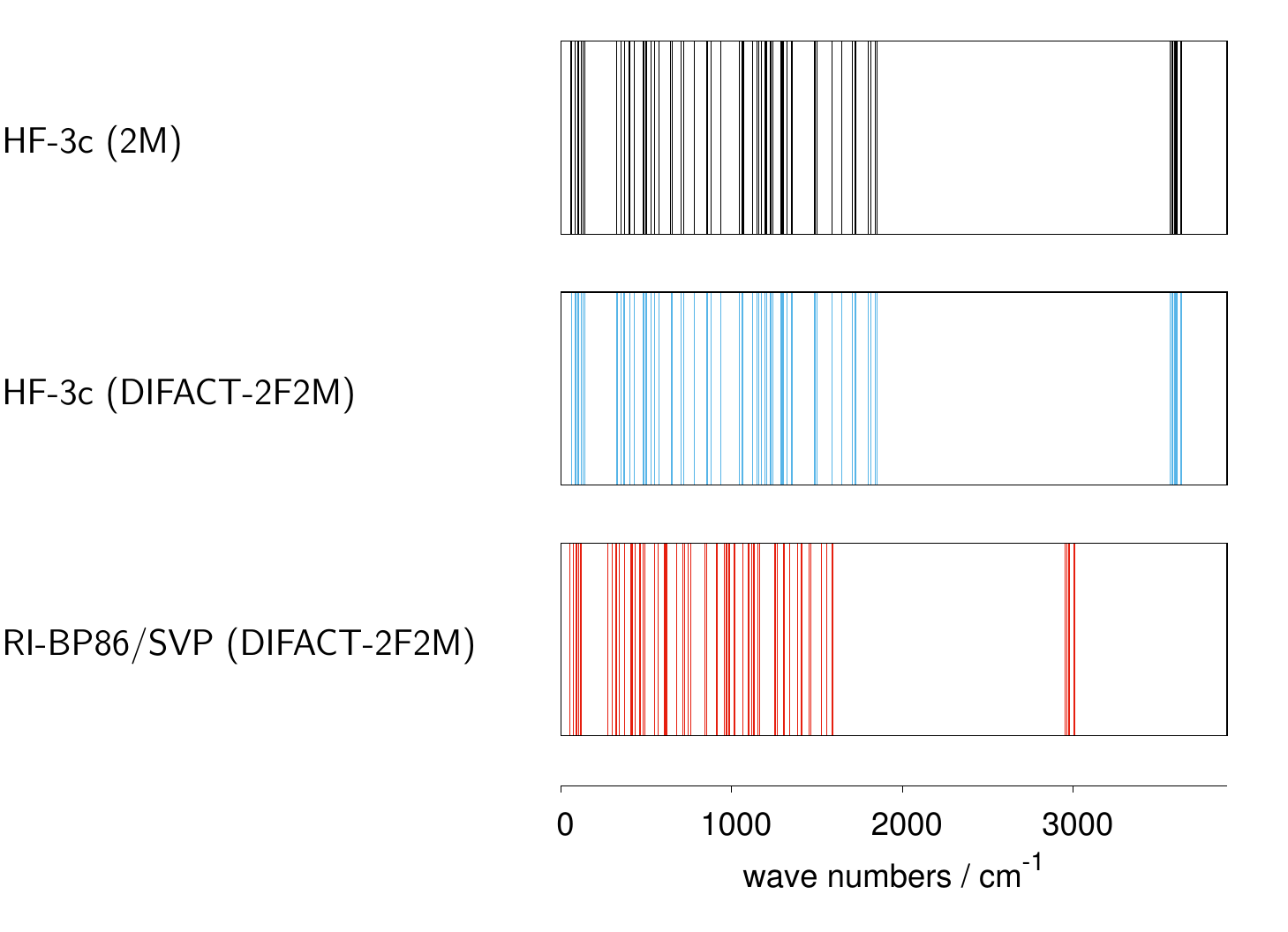}\\
\caption{
Stick spectra of VSCF fundamental excitation energies  in the applied FALCON coordinates 
 for HF-3c and RI-BP86/SVP in selected representations of the PES, i.e., 
 2M and DIFACT-2F2M.
\label{fig:bp86}
}
\end{figure}
It is striking that the differences between the different PES generation schemes using HF-3c is hardly 
 visible, whereas the difference due to the different electronic structure method is rather significant.
It should also be mentioned 
  that the frequently applied frequency scaling for the two methods compared here 
 is rather different: In Ref.~\onlinecite{sure2013}, a frequency scaling factor of 0.86 is applied for HF-3c frequencies, 
 whereas in a recent study fitting different frequency scaling factors\cite{kesh2015}, 
 a frequency scaling factor of 1.03 for BP86/def2-SVP is advertised. 
Note that def2-SVP is equivalent to SVP for carbon and hydrogen atoms, i.e., 
 the only atoms contained in the oligo-phenyl examples. 
This means we cannot really expect the VSCF results for PESs obtained with these
 two electronic structure methods to be similar.  
Nevertheless, the strong effect of the choice of the electronic structure method compared 
 to the errors introduced by the double incremental expansion is remarkable. 
 With the DIF and DIFACT schemes methods one has new methods for finding
 good compromises between accuracy and efficiency. 
Thus, a double incremental PES expansion with a more expensive electronic structure method
 may be a more promising route to efficient generation of reliable PESs for 
 vibrational structure calculations for larger systems,
 than to compromise the potential by using a lower-level electronic structure method,
 or avoiding higher-level mode couplings.

\section{Summary, Conclusions and Outlook \label{sec:conc}}

We have presented a double incremental expansion for the generation of PES
  for vibrational structure calculations
  combining the $n$-mode expansion of the PES with the many-body expansion of the 
  underlying evaluations of the electronic energies.
We have shown that this type of double incremental expansion of the PES is 
 particularly beneficial when combined with semi-local coordinates to span the 
 vibrational space.
This leads to a  significant reduction of the number of required SPCs. 
For this reason, we use the recently introduced rectilinear and semi-local 
 FALCON coordinates\cite{koni2016} in combination with the double incremental expansion 
 leading to the double incremental scheme in FALCON coordinates (DIF).
Even in these coordinates, the number of required modes to be considered for the PES 
 generation for fragment combinations often exceeds the number of vibrational 
 degrees of freedom in the respective fragment combination. 
This means, that the PESs of the fragment 
 combinations are spanned in a redundant basis 
 and the explicit calculation of grid points in this redundant basis leads to similar 
 overall scaling behavior as in delocalized modes. 
This scaling behavior of the DIF scheme is caused by a limited number of modes.   
These are so-called inter-connecting modes, i.e., 
 those modes that describe the relative motion of the fragments. 
In an alternative scheme, we can express these motions by auxiliary modes. 
In this scheme,
 the generation of the PES for every fragment combination 
  only considers the number of vibrational 
 degrees of freedom in the respective fragment combination plus three modes corresponding 
 to rotations. 
In this way, the number of required SPCs per fragment combination 
 is independent of the total number of fragments and \chan{an} overall low scaling 
 behavior can be achieved. 
This is, however, achieved at the cost of introducing an additional 
 error in the subsequent transformation of the fragment combinations' PESs 
 to the common coordinates.
We stress that the  scaling analysis does not only consider the number of required
 SPCs, but also takes the cost per SPC into account.
No assumptions of a particular scaling of the individual \chan{SPCs} with 
 system size has been made. 
Our approach differs from previous low scaling approaches for the generation 
 of PESs\cite{hans2010,rich2014}, where ``linear scaling'' refers to the 
 number of required SPCs but not to the accumulated computational cost 
 of these SPCs.

Numerical examples on tetra- and hexa-phenyls show quick convergence of the 
 VSCF fundamental excitation energies for PESs constructed with increasing 
 level in fragment combination. 
Indeed the DIF-3F2M results resemble the 2M results almost perfectly in the tested cases.  
Using auxiliary coordinates, i.e., the corresponding DIFACT scheme, leads to additional 
 RMSDs of below 1~{\rcm} with maximum 
 absolute deviations of less than 5~{\rcm}.
The larger transformation errors are, in these examples,
  restricted to very few fundamental 
 excitation energies that correspond to inter-connecting modes.
The overall modest size of these errors can, at least partly, 
 be related to relative close 
 resemblance of the overall FALCON modes to the auxiliary modes of the same type.
We have illustrated that this error lies clearly below the difference in fundamental 
 excitation energies obtained with different electronic structure methods realistically applicable 
 in the generation of PESs for systems containing tens of atoms or more.  
This suggests that it may be more relevant to choose a more accurate and thereby a computationally 
 more demanding electronic structure method, but at low order in the double incremental expansion 
 to yield the most beneficial balance of computational cost and accuracy.
We foresee that the computationally cheapest double incremental expansion, i.e., DIFACT-2F2M
 may be sufficient in many cases and will enable reliable generation of PESs
 for wave function-based calculations of vibrational spectra, for system sizes, that 
 are currently clearly out of reach.  
 
 \chan{The considered test systems are perfect examples for which our methods are appropriate, 
 as chain-like systems with modest long-range electrostatic interactions and capping adequately handled by     
 hydrogens. 
In addition, the level of the electronic structure description applied was rudimentary.  
 For many other complex systems there may be a slower decay in importance
 with distance between fragments
 as well as more advanced capping schemes may be preferred. 
It is an essential ingredient of the method that only changes
 in energy relative to a reference structure are considered. 
Thus, even if the caps do not cancel for
 a given structure, their contribution may cancel out in the contributions to the PES.
This is likely to simplify the use of other capping schemes as well,
  as long  as rigid caps are used.
 Nevertheless, it must be carefully considered 
 which steps need to be included 
 in applying the method to complex systems with advanced cappings. Likewise, the rate of decay with 
 distance for the fragment couplings will vary between systems, and potentially also between different computational methods.
 For example, the accuracy of the fragment many-body expansion has been suggested to
 depend on the character of the applied basis set and potentially involves a basis set superposition error.\cite{ouyang_trouble_2014,yuan_are_2016,lao_understanding_2016}  }

\chan{In addition to the above-mentioned examples much other work is needed
 extending the methodology and acquiring experience before the double incremental approach can become a widely used standard procedure. }
 The present work focuses on PESs and the extension to other 
 property surfaces such as dipole surfaces, is needed to calculate infrared spectra.
Generally, the extension to size extensive, additive properties will be straightforward.  
Another point of possible improvement concerns the currently applied transformation 
 algorithm. 
Other more elaborate schemes, for instance using interpolation techniques may reduce the 
 transformation error and thereby extend the applicability of the very efficient 
 DIFACT scheme. 
 For efficiency and {increasing the} black box nature, 
 it is also important to integrate the double incremental approach
 with automatic choice of required grid points (as, for instance, done in the 
 adaptive density-guided approach (ADGA)\cite{spar2009}).
Multi-level\cite{rauh2004,yagi2007,rauh2008,spar2009b,spar2010,meie2013} 
 and screening approaches\cite{beno2004,rauh2004,beno2006,pele2008,beno2008,seid2009b} as well as 
 various interpolation and extrapolation schemes\cite{rauh2004,mati2009,spar2010} can 
 likewise be integrated with the double incremental approach.
To combine these ideas in a coherent framework will be topics for future research. 
 Finally, the methodology developed here should also be well applicable using 
 curvilinear internal coordinates,
 and at least some aspects of the theory will be simpler, such as the definition of 
 locality, and rigidity under relative rotations etc.
 With the nontrivial  construction of the appropriate kinetic energy operator for such coordinates
 becoming more automatic through recent research\cite{lauv2002,ndon2012}, the development of a double incremental scheme in curvilinear coordinates
 is a challenging but not unrealistic topic of future research.

All in all, the presented double incremental scheme for the generation of PESs 
 pushes size limitation for PES generation with quantum chemical 
 evaluation of the 
 electronic energy points to significantly larger sizes compared to the conventional $n$-mode expansion. 
Combined with other schemes to gain more efficiency, this method will be an important 
 step towards reliable PESs and thereby towards vibrational structure 
 calculations for sizable systems, including covalently bound molecules.
We expect our approach to be a valuable complement to 
 vibrational wave function approaches for bio-molecular vibrational 
 spectroscopy\cite{ades2005,pane2014,roy2015,roy2016} as well as  molecular vibrations in heterogeneous environments, 
 such as molecules on metal surfaces\cite{ulus2011,chul2013,beno2015}.

\section*{Acknowledgments}
We are grateful to Ian H.\ Godtliebsen for continuous support 
 in the development of the MidasCpp code as well 
 as Emil Lund Klinting, Mads B{\o}ttger Hansen, Diana Madsen, Niels Kristian 
 Madsen, and Gunnar Schmitz for careful reading of this manuscript.
C.K.\ acknowledges funding by a Feodor--Lynen research fellowship from the Alexander von Humboldt 
 Foundation as well as a post-doctoral grant from the Carlsberg Foundation. 
 O.C. acknowledges support from the Lundbeck Foundation, 
 the Danish e-infrastructure Cooperation (DeiC), and the Danish Council for Independent Research through a Sapere Aude III grant (DFF -- 4002-00015).

\bibliographystyle{aipnum4-1}

\begin{appendix}

\chan{
\section{Derivation of Eq.\ (\ref{eq:coef}) \label{sec:app_A}}

For calculating  the coefficients $p_{{\bf c}_{l}}^{\rm VCR}$ of the 
cut function $F_{{\bf c}_{l}}$ in an 
 incremental expansion for a variable combination range VCR 
 according to Eq.~(\ref{eq:overall_exp}) we need to consider all 
 variable combinations ${\bf c}_{s}$ in VCR  
 that contain ${\bf c}_{l}$ or are equal to ${\bf c}_{l}$, i.e. ${\bf c}_{s} \supseteq {\bf c}_{l}$.
Since we require the VCR to be closed on forming subsets, the
 contribution of a superset ${\bf c}_{s}$ to the coefficient of $ {\bf c}_{l}$ 
 only depends on the order of the involved variable 
 combinations, i.e., the values of $l$ and $s$.
They can be summarized as

\begin{equation}
\tilde {p}^{s}_{l}
=
\left\{
\begin{split}
0 \quad   \forall & s < l \\
1 \quad  \forall & s = l \\
-\sum_{l'=l}^{s-1} 
\begin{pmatrix}
s-l \\ l' -l 
\end{pmatrix}
\tilde{p}^{l'}_l
\quad
   \forall & s > l \\
\end{split}
\right.
.
\end{equation}
For variable combinations  ${\bf c}_{s}$ that are smaller than ${\bf c}_{l}$, 
 ${\bf c}_{l}$ cannot be a subset of or equal to ${\bf c}_{s}$ and we obtain
 $\tilde {p}^{s}_{l} = 0 \quad \forall   s < l $.
For $s = l$, $F^{{\bf c}_{l}}=F^{{\bf c}_{s}}$ occurs only once 
 in $\bar{F}^{{\bf c}_{s}}$ [see also Eq.~(\ref{eq:incr_bar_combi})],
  so that $\tilde {p}^{s}_{l} = 1 \quad \forall   s = l $.
If $ s > l$, $F^{{\bf c}_{l}}$ can contribute to several terms  
 in $\bar{F}^{{\bf c}_{s}}$ [Eq.~(\ref{eq:incr_bar_combi})].
It is contained in 
 the bar potentials for 
 all variable combinations that are supersets of  ${\bf c}_{l}$. 
Since these terms are correcting terms in the bar potential $\bar{F}^{{\bf c}_{s}}$, 
 they have to be subtracted, which  
 explains the minus sign of the contribution. 
Each term occurs $\begin{pmatrix} s-l \\ l'-l \end{pmatrix}$ times, which 
 is the number of subsets of  ${\bf c}_{s}$ that contain ${\bf c}_{l}$ and have the 
 order $l'$.
Let us, for example look at the coefficient of the variable combination 
 $\{A\}$. 
Its coefficient in $\bar{F}^{\{A\}}(\{x_A\})$ is trivially 
\begin{align}
\tilde{p}^1_1 = 1
\end{align}
In 
\begin{align}
\bar{F}^{\{A,B \}}(\{x_A,x_B\}) = {F}^{\{A,B \}}(\{x_A,x_B\}) - \bar{F}^{\{A\}}(\{x_A\}) - \bar{F}^{\{B\}}(\{x_B\})
\end{align}
${F}^{\{A\}}(\{x_A\})$ has the coefficients
\begin{align}
\tilde{p}^2_1 = -\tilde{p}^1_1 = -1
\end{align}
and to 
\begin{align}
\bar{F}^{\{A,B,C\}}(\{x_A,x_B,x_C\}) = & {F}^{\{A,B ,C\}}(\{x_A,x_B,x_C\}) 
 \notag \\  
 & -
 \bar{F}^{\{A,B\}}(\{x_A,x_B\}) 
 - \bar{F}^{\{A,C\}}(\{x_A,x_C\})
 - \bar{F}^{\{B,C\}}(\{x_B,x_C\}) 
 \notag \\
& -
 \bar{F}^{\{A\}}(\{x_A\}) 
- \bar{F}^{\{B\}}(\{x_B\})
- \bar{F}^{\{C\}}(\{x_C\})
\end{align}
it contributes with 
\begin{align}
\tilde{p}^3_1 = - \left( 
 \begin{pmatrix} 
  2 \\ 0
 \end{pmatrix} 
 \tilde{p}^1_1 
+
 \begin{pmatrix} 
  2 \\ 1
 \end{pmatrix} 
 \tilde{p}^2_1 
 \right)
 = 
- \left( 1\cdot 1 + 2 \cdot (-1) \right)
=
 1.
\end{align}
These observations can easily be 
 generalized to 
\begin{align}
\tilde{p}_l^l = 1  , \quad 
\tilde{p}_l^{l+1} = -1,  {\rm and} \quad 
\tilde{p}_l^{l+2} = 1 .
\end{align}
This suggests that 
\begin{align}
\tilde{p}^s_l  = (-1)^{s-l}.
\label{eq:suggestion}
\end{align}
We will prove in the following by induction that, 
 if Eq.~(\ref{eq:suggestion})  holds for $l<s<(n+1)$ (as shown 
 above for $l< s < l+3$), it is also true for $s=n+1$ and thereby for any 
 $s>l$, $s,l \in \mathbb{N}$.
Therefore we consider
\begin{align} 
\tilde{p}^{n+1}_l 
 & = 
 - \sum_{l'=l}^n 
\begin{pmatrix}
n-l+1 \\ l' -l 
\end{pmatrix}
\tilde{p}^{l'}_l
= 
 - \sum_{l'=l}^n 
\begin{pmatrix}
n-l+1 \\ l' -l 
\end{pmatrix}
(-1)^{l'-l} 
\notag \\
& = 
 - \sum_{l'=l}^{n+1} 
\begin{pmatrix}
n-l+1 \\ l' -l 
\end{pmatrix}
(-1)^{l'-l} 
 + 
\begin{pmatrix}
n-l+1 \\ n -l +1
\end{pmatrix}
(-1)^{n+1-l} 
\notag \\
& = (-1)^{n+1-l} ,
\end{align} 
where we have used $\sum_{l''=0}^{n-l+1} 
\begin{pmatrix}
n-l+1 \\ l'' 
\end{pmatrix}
(-1)^{l''}
= 0
$
with $l'' = l' -l$.
Thereby we have proven that the coefficient of a variable combination 
 ${\bf c}_l$ to the bar function of a fragment combination 
 ${\bf c}_s \supseteq {\bf c}_l$ 
 is given by $\tilde{p}^s_{l}= (-1)^{s-l}$.

With this we can obtain the coefficients $p_{{\mbf c}_{l'}}^{\rm VCR}$ used 
in Eq.~(14) as the sum of the coefficients $\tilde{p}^l_{l'}$ arising from all higher-order variable 
combinations ${\mbf c}_l$ in VCR that are supersets of  ${\mbf c}_{l'}$ as
\begin{align}
p_{{\mbf c}_{l'}}^{\rm VCR} = 
\sum_{{\mbf c}_l \in {\rm VCR}; {\mbf c}_{l} \supseteq {\mbf c}_{l'}}
\tilde{p}^l_{l'}
= 
\sum_{{\mbf c}_l \in {\rm VCR}; {\mbf c}_{l} \supseteq {\mbf c}_{l'}}
(-1)^{l-l'}. 
\end{align}
Thereby, we have derived Eq.~(15).
}

\section{Effective spatially restricted fragment combination range with only neighbor couplings for a chain-like system \label{app:chain-eff}}
Following Eq.~(\ref{eq:coef}), the coefficient of a particular fragment combination 
 ${\mbf f}_{k}$ in the overall incremental expansion of the energy,  $p_{{\mbf f}_{k}}^{\rm FCR}$, 
 is dependent on the FCR. 
Assume, we have a chain-like system and a spatially restricted FCR with 
 only neighbor couplings and a highest fragment-combination order of $l$.
When adding a fragment combination of size $l+1$, we need additionally to add all subsets that 
 are not yet included. 
For instance, the inclusion of a fragment combination $\{A,B,C,D\}$ to the ${\rm FCR^{3F,NB}}$ in
 Eq.~(\ref{eq:FCR_spatialexample}), has to be accompanied by the addition of 
 the fragment combinations $\{A,D\}$, $\{A,B,D\}$, and $\{A,C,D\}$.
This ensures that the resulting FCR is closed on forming subsets.
All these additionally required fragment combinations contain the outer-most fragments of the 
 new fragment combination, i.e.\chan{,} $A$ and $D$ in the present example. 
The number of required \chan{additional} fragment combinations for a certain fragment combination level $k$ is, hence,
 determined by the number of additional permutations possible.
It can be obtained by $\begin{pmatrix} l'-2 \\ k-2 \end{pmatrix}$, where $l'=l+1$ is the level of 
 the newly added fragment combination ${\mbf f}_{l'}$. 

The extension of the FCR by ${\mbf f}_{l'}$, ensuring that it is closed under 
 forming subsets, may \chan{hence} affect the coefficients in the expansion  analogous to Eq.~(\ref{eq:overall_exp}) 
 for all fragment combinations that are subsets of or equal to ${\mbf f}_{l'}$.
We denote the change in coefficients for a fragment combination ${\mbf f}_{k}$ 
 by $\Delta p_{{\mbf f}_{k}}$.
First we consider the fragment combinations \chan{${\bf f}_k$} 
 that are subsets of or equal to 
 the added fragment combination ${\mbf f}_{l'}$ itself but not of any 
 other {\chan{\it added}} fragment combination.
This is the only the case for ${\mbf f}_{l'}$ itself and the 
  connected mode combinations of size $k=l'-1$ ($\{A,B,C\}$, and $\{B,C,D\}$ in the 
 above example) as well as the connected fragment combination 
  of size  $k=l'-2$ that does not contain the outer fragments ($\{B,C\}$ in the 
 above example).
In these cases, we obtain $\Delta p_{{\mbf f}_{k}}=(-1)^{l'-k}$, which is the 
 contribution due to the largest added fragment combination (${\mbf f}_{l'}$) alone.
Smaller connected fragment combinations will be contained 
 in the additionally added fragment combinations.
In these cases and \chan{for the} unconnected 
 fragment combinations, we have to ``count'' how often the particular fragment combination is contained in 
 the added fragment combinations. 
In these cases we obtain, when adding a fragment combination of level $l'$ (${\mbf f}_{l'}$), 
 a change of overall coefficient 
 for a fragment combination ${\mbf f}_{k}$, where ${\mbf f}_{k} \subset {\mbf f}_{l'}$, 
 by applying Eq.~(\ref{eq:coef}),
\begin{equation} 
\Delta p_{{\mbf f}_{k}} 
= 
(-1)^{l'-k}
+
\sum_{i=k+2-o}^{l'-1} 
\begin{pmatrix} l'-k-2+o \\ i-k-2+o \end{pmatrix}
(-1)^{i-k}
\label{eq:chain-coeff-general}
,
\end{equation} 
where $o$  is the number of outer-most fragments of ${\mbf f}_{l'}$ contained 
 in ${\mbf f}_{k}$ , i.e.,  \chan{$o\in \{0,1,2\}$}, so that  
 $k \ge o$ and  $ k \le l'$.  
The first term in Eq.~(\ref{eq:chain-coeff-general}) arises from the added fragment 
 combination ${\mbf f}_{l'}$  and
 the sum collects all contributions due to the additional 
 fragment combinations added to ensure the FCR to be closed on forming subsets. 
The binomial coefficient is obtained by consideration on how many permutations
 need to be considered when fixing the $k$ fragments in ${\bf f}_k$ \chan{and 
 the $2-o$ outer-most fragments} that are not contained in ${\bf f}_k$.  
As an example, we consider $\Delta p_{{\mbf f}_{k}}$ for ${\mbf f}_{k}= \{A,B\}$, 
 with $k=2$, in the 
 above example, where $\{A,B,C,D\}$ was added to the ${\rm FCR^{3F,NB}}$ in
 Eq.~(\ref{eq:FCR_spatialexample}).
The size of the largest added fragment combination is, hence, $l' = 4$.  
We see that $\{A,B\}$ contains one outer-most fragment, namely A, so that $o=1$
 and obtain
\begin{equation}
\Delta p_{{\mbf f}_{k}} 
= 
(-1)^{4-2}
+
\sum_{i=3}^{3} 
\begin{pmatrix} 1 \\ 0 \end{pmatrix}
(-1)^{3-2}= 1-1 =0
.
\end{equation} 
The first contribution, here is that of the added fragment combination $\{A,B,C,D\}$, 
 and the second one that of $\{A,B,D\}$.

In the special case that $k=0 \land o=0 \land l' \le 2 $, the sum in Eq.~(\ref{eq:chain-coeff-general})
 does not contribute and we obtain
\begin{equation}
\Delta p_{{\mbf f}_{k}} = (-1)^{l'} \qquad \forall (k=0 \land o=0 \land l' \le 2).
\end{equation}
In the other cases, we combine the sum with the contribution of the largest 
 added fragment of size $l'$\chan{ in Eq.~(\ref{eq:chain-coeff-general})}, so that  
\begin{equation}
\Delta p_{{\mbf f}_{k}} = 
\sum_{i=k+2-o}^{l'} 
\begin{pmatrix} l'-k-2+o \\ i-k-2+o \end{pmatrix}
(-1)^{i-k}
=
\sum_{i'=0}^{l''}
\begin{pmatrix} l'' \\ i' \end{pmatrix}
(-1)^{i'}
= 0
 \qquad \forall \neg (k=0 \land o=0 \land l' \le 2),
\end{equation}
where $l''=l'-k-2+o > 0$ and $i'= i-k-2+o$.

This means that the only fragment combinations that are subject to a non-zero change in the
 coefficients in this scheme are those of the added fragment combination of size $l'$ 
 itself as well as the
 connected fragment combination of one order lower than the added one ($k=l'-1$), 
 the connected fragment combination with $k=l'-2$ that does not contain an 
 outer-most fragment, and the zero set, if $l'\le 2$. 
\chan{The} contributions of all other cases cancel exactly. 
A consequence of this is that, for a chain-like molecule using a spatially restricted FCR with 
 only neighbor couplings, 
 there are no contributions from unconnected fragment combinations.
This holds true for all levels of fragment combinations.

Building up the spatially restricted FCR with only neighbor couplings for a chain-like system is 
 exercised in Table~\ref{tab:chain-coef}.
\begin{table}
\caption{
Coefficients $p^{f{\rm F}}$ and change of coefficient $\Delta p^{(f+1){\rm F} - f{\rm F}}$
 when adding the next $f+1$-level of fragment combinations to the   
 spatially restricted FCR with only neighbor couplings 
 of a chain-like system with $N$ fragments for different fragment combinations (FC) 
 and fragment combination order $f$.
\label{tab:chain-coef} 
}
\begin{tabular}{lcccccccccc}
\hline
\hline
FC                     & $p^{\rm 1F}$ & $\Delta p^{\rm 2F-1F}$ 
 & $p^{\rm 2F}$ & $\Delta p^{\rm 3F-2F}$ 
 & $p^{\rm 3F}$ & $\Delta p^{\rm 4F-3F}$ 
 & $p^{\rm 4F}$ 
 & $\dots$ \\ 
\hline                  
$\{\}$                 &  $(N-1)$       & $(N-1)\cdot (-1)$ & 0 & 0 & 0 & 0 & 0 & $\dots$ \\
\hline                  
$\{F1\},\{FN\}$        &  1           & $1 \cdot (-1)$    & 0 & 0 & 0 & 0 & 0 &  $\dots$\\
$\{F2\}, \dots\{F(N-1)\}$ &  1        & $2 \cdot (-1)$    & $-$1 &  $1 \cdot (1)$   & 0 & 0 & 0 & $\dots$\\
\hline                  
$\{F1,F2\},\{F(N-1),FN\}$        & --- & 1 &  1           & $1 \cdot (-1)$    & 0 & 0 & 0 & $\dots$\\
$\{F2,F3\}, \dots ,\{F(N-2),F(N-1)\}$ & --- & 1 &  1        & $2 \cdot (-1)$    & $-$1 &  $1 \cdot (1)$   & 0 & $\dots$\\
\hline                  
$\{F1,F2,F3\},\{F(N-2),F(N-1),FN\}$        & --- & --- & --- & 1 &  1           & $1 \cdot (-1)$    & 0  & $\dots$\\
$\{F2,F3,F4\}, \dots ,\{F(N-3),F(N-2),F(N-1)\}$ & --- & --- & --- & 1 &  1        & $2 \cdot (-1)$    & $-$1 &  \\
\hline
\hline
\end{tabular}
\end{table}
We see that the first time a fragment combination of size $l$ is introduced it 
 has a coefficient of $1$. 
The inclusion of the next-higher $(l+1)$-level of fragment combinations then leads to two (or one in case of outer fragment combination) 
 contributions of $-1$.
This is due to the fact, that these fragment combinations are connected fragment combinations of level $l$
 that do not contain both outer fragments of a fragment combination of level $l+1$. 
In case these $l$-level fragment combinations do not contain any of the
 outer-most fragments of the entire chain-like system, they 
 are contained in two of the connected fragment combinations of size $l+1$. 
This leads to a coefficient of $1+2\cdot(-1)= -1$. 
If it contains one outer-most fragment of the chain, it is only contained in one of the added 
  fragment combinations of size $l+1$, leading to a coefficient of  $1+1\cdot(-1)= 0$. 
Adding the next level $(l+2)$, can only change the coefficients by $1$ for those 
 $l$-level fragment combinations 
 that are a contained in an $(l+2)$-level fragment combination, but do not contain any outer fragment of the latter  
 fragment combination. 
This case occurs exactly once for each $l$-level fragment combination that does not contain 
 an outer-most fragment of the chain. 
The coefficients of these in the $(l+2)$F expansion are therefore zero. 
As shown above, the addition of higher-order fragment combinations in this scheme will not have any influence on the 
 coefficients, so that they will be zero for all these higher FCRs.
The results can be summarized for $f>1$ as 
\begin{equation}   
p_{{\mbf f}_k}^{f{\rm F, NB}}
=
\left\{
\begin{split}
& 1 & \forall  k=f  \\
& -1 & \forall k = (f-1) \land {\mbf f}_k\ {\rm connected} \   \land {\mbf f}_k \cap \{F1,FN\} = \emptyset \\ 
& 0 & {\rm else}
\end{split}
\right.
,
\end{equation}   
\chan{where $\{F1,FN\}$ is the set of the two outer-most fragments in the chain-like system.}
This means that the only fragment combinations that contribute to a 
 spatially restricted $f$-level FCR in a
 chain like system are those of level $f$ and $f-1$, where the latter only have a non-zero contribution, 
 in case they do not contain an outer-most fragment of the chain.

\section{Coordinate transformation of polynomial PES representations \label{app:trans}}

In the current work, we apply transformation between polynomial representations of PESs for 
 certain fragment combinations (FCs).
The starting auxiliary PES is represented in a polynomial sum-over-product form 
 in auxiliary coordinates as,
\begin{align}
V^{\rm aux}_{FC} (\{q^{\rm aux}_{\rm FC}\})
 = \sum_{\mbf{t}} (a_{\rm FC}^{\rm aux})_{\mbf{t}} \prod_{m \in \mbf{t}} q_{m}^{\rm aux},
\end{align}
where $\mbf{t}$ denotes a particular term in the potential. 
For instance, $\mbf{t} = \{q_1,q_1,q_1,q_3,q_3,q_5\}$ corresponds to the potential term 
 $q_1^3 \cdot q_3^2 \cdot q_5$.
Due to permutational symmetry, the same potential term is also obtained for other   
 $\mbf{t}$s.
This is the case, for instance, for  $\mbf{t}_2 = \{q_1,q_1,q_5,q_3,q_3,q_1\}$
 and  $\mbf{t}_3 = \{q_1,q_1,q_5,q_3,q_1,q_3\}$ and we have 
 $(a_{\rm FC}^{\rm aux})_{\mbf{t}} =  (a_{\rm FC}^{\rm aux})_{\mbf{t}_2} =  (a_{\rm FC}^{\rm aux})_{\mbf{t}_3}$.
We can, hence, collect these terms and express the PES as 
\begin{align}
V^{\rm aux}_{FC} (\{q^{\rm aux}_{\rm FC}\})
 = \sum_{\mbf{s}} (c_{\rm FC}^{\rm aux})_{\mbf{s}} \prod_{m \in \mbf{s}} (q_{m}^{\rm aux})^{e_m^{\mbf{s}}},
\end{align}
where ${\mbf{s}}$ contains the modes in this term as well as the power $e_m^{\mbf{s}}$ with which it is 
 represented in the potential term.
For $\mbf{t}$ and $\mbf{s}$ corresponding to the same term, we have
\begin{align}
 (c_{\rm FC}^{\rm aux})_{\mbf{s}} = u_{\mbf{s}} (a_{\rm FC}^{\rm aux})_{\mbf{t}},
\end{align}
where $u_{\mbf{s}} = \frac{(\sum_{m \in \mbf{s}}e^\mbf{s}_m)!}{\prod e^\mbf{s}_m!}$ is the number of
 possible permutations\chan{, i.e., the number of corresponding $\mbf{t}s$}. 

Also the final PES in common FALCON coordinates can be written in both ways, i.e., as 
\begin{align}
V^{\rm F}_{FC} (\{q^{\rm F}_{\rm FC}\})
 = \sum_{\mbf{t}} (a_{\rm FC}^{\rm F})_{\mbf{t}} \prod_{m \in \mbf{t}} q_{m}^{\rm F},
\end{align}
or 
\begin{align}
V^{\rm F}_{FC} (\{q^{\rm F}_{\rm FC}\})
 = \sum_{\mbf{s}} (c_{\rm FC}^{\rm F})_{\mbf{s}} \prod_{m \in \mbf{s}} (q_{m}^{\rm F})^{e_m^{\mbf{s}}},
\end{align}
where  $q^{\rm F}_{\rm FC} $ refers to that part of the common FALCON 
 coordinates that displace the atoms in the respective fragment combination\chan{ (FC)}. 

The transformation from the PES in auxiliary coordinates $\{q\}^{\rm aux}_{\rm FC}$ 
 to  $\{q\}^{\rm F}_{\rm FC}$, 
 can be performed via subsequent addition of terms for the $\{q\}^{\rm F}_{\rm FC}$ coordinates 
 as linear combinations of PES terms containing auxiliary coordinates.
The weights in this linear combination are given by the elements of the transformation matrix $\mbf{A}_{\rm FC}$.
We can, for instance obtain,
\begin{align}
(a^{\rm F}_{\rm FC})_{m_1,m_2, \dots,\tilde{m}_f, \dots,  m_n} 
 = \sum_{a \in \mbf{m}^{\rm aux}} 
 (\mbf{A}_{\rm FC})_{fa}
(a^{\rm aux}_{\rm FC})_{m_1,m_2, \dots,m_a, \dots, m_n},
\end{align}
where $\tilde{m}_f$ denotes the \emph{new} common FALCON coordinate and $a$ any auxiliary 
 coordinate.
When collecting the terms similar to above, we have to take the permutational symmetry into account,
\begin{align}
(c^{\rm F}_{\rm FC})_{\tilde{\mbf{s}}(f)} 
 =  u_{\tilde{\mbf{s}}(f)}\sum_{a \in \mbf{m}^{\rm aux}} 
 (\mbf{A}_{\rm FC})_{fa}
   u_{\tilde{\mbf{s}}(a)}^{-1}
(c^{\rm aux}_{\rm FC})_{\tilde{\mbf{s}}(a)},
\end{align}
where $\tilde{\mbf{s}}(f)$ indicates, similar to above, that one coordinate of the 
 original PES representation in auxiliary coordinates
 has been replaced by a new, common FALCON coordinate and 
 in $\tilde{\mbf{s}}(a)$ the same spot is occupied by the 
 auxiliary coordinate $a$.
$\mbf{m}^{\rm aux}$ is the full set of all auxiliary coordinates.
In this way we can add term after term for the new coordinates. 
The terms with higher order $e_f$ in the new coordinates are subsequently obtained 
 from the $e_f-1$ terms that contain auxiliary coordinates. 

In the transformation algorithm, the potential terms containing auxiliary coordinates are 
 kept and re-used in subsequent addition of coordinates.
Despite the fact that we transform \chan{an} $n$-mode representation in auxiliary coordinates to 
 \chan{an} $n$-mode 
 representation in common FALCON coordinates, coefficients for ($n+1$)-mode combinations with at least 
 one auxiliary coordinate occur temporarily.
We do not consider any terms of higher-order in mode combination, since they only contribute to    
  ($n+1$)- or higher-order mode combinations in the final PES representation, which we neglect.
The \chan{described} subsequent addition of coordinates to a PES in two-mode representation in this way 
 scales with $M_{\rm aux}^3$  and $M_{\rm F}^2$, where  $M_{\rm aux}$ is the number of 
 auxiliary coordinates and $M_{\rm F}$ that of common FALCON coordinates to be considered 
 in the respective fragment combination.
At the end of the transformation, all contributions from auxiliary coordinates are removed from the 
 representation of the PES.
\end{appendix}

\end{document}